\documentclass[11pt,a4paper]{article}
\usepackage{jheppub}
\usepackage{amssymb,amsmath,amsfonts,mathrsfs}
\usepackage{graphicx}
\usepackage[dvipsnames]{xcolor}
\usepackage{verbatim}
\usepackage{epsfig}
\usepackage{color}
\usepackage{multirow}
\usepackage{comment}
\usepackage{datetime}
\usepackage{slashed}
\usepackage{ulem}
\usepackage{framed}
\usepackage{longtable}
\usepackage[mathscr]{euscript}
\usepackage{cancel}
\usepackage{tensor}
 \usepackage{mathtools}
  \usepackage{caption}
  \usepackage{subcaption}
 \usepackage[multiple]{footmisc}
 \usepackage{pgfplots}
 \usepackage{natbib}
\usepackage[applemac]{inputenc}
\usepackage{tikz}
\usepackage{placeins}
\usetikzlibrary{decorations.markings}

\newcommand{\nn}{\nonumber}

\def\be{\begin{equation}}
\def\ee{\end{equation}}
\def\bea{\begin{align}}
\def\eea{\end{align}}

\def\a{\alpha}      
      \def\q{\theta}
  \def\k{\kappa}    \def\m{\mu}
\def\n{\nu}

\title{Scalar Quasinormal modes in Reissner--Nordstr\"om black holes: implications for Weak Gravity Conjecture}

\author{Giorgio Di Russo$^a$,}

\author{Anna Tokareva$^{a,b,c}$}

\affiliation{$^a$School of Fundamental Physics and Mathematical Sciences, Hangzhou Institute for Advanced Study, UCAS, Hangzhou 310024, China}
\affiliation{$^b$Department of Physics, Blackett Laboratory, Imperial College London, SW7 2AZ London, UK}
\affiliation{$^c$International Centre for Theoretical Physics Asia-Pacific, Beijing/Hangzhou, China}

\abstract{Microscopic charged black holes can provide possibilities to test the consistency of the effective field theory (EFT) corrections to Einstein-Maxwell theory. A particularly interesting result is fixing the sign of a certain combination of EFT couplings from the requirement that all charged black holes should be able to evaporate (Weak Gravity Conjecture). In our work, we analysed the EFT corrections to a set of zero-damping quasinormal modes (QNMs) of the scalar wave probe in a nearly extremal Reissner-Nordstr\"om black hole. We review the duality of this setup to the problem of the quantum Seiberg-Witten curve of $N=2$ Super-Yang-Mills theory with three flavors. We provide an analytic result for the EFT corrections to the QNMs obtained from the quantization condition imposed on the Seiberg-Witten cycle. Our main result is that the causality requirement of the gravitational theory formulated for the QNMs translates to the same condition on EFT couplings as the one appearing in the Weak Gravity Conjecture.}

\begin{document}
\maketitle
\flushbottom 
\section{Introduction}

Properties of charged black hole solutions are of great interest, as they can provide an additional consistency probe for the theory describing gravity. The existence of critically charged black holes (without naked singularities) is important for the very possibility of a black hole with arbitrary mass and charge to evaporate. The corresponding statement is known as the Weak Gravity Conjecture (WGC) \cite{Arkani-Hamed:2006emk}, stating that in any theory with a $U(1)$ gauge field and gravity, there must be a state in the spectrum with a charge-to-mass ratio $Q/M > 1$ (in Planck units). Originally proposed to ensure the decay of extremal black holes and avoid remnants, the WGC has since been generalized to non-Abelian gauge theories, higher-form symmetries, and scalar fields~\cite{Rudelius:2015xta,Montero:2018fns}. The importance of the WGC lies in its role as a condition for an EFT to admit a self-consistent UV completion in quantum gravity. It ensures that extremal black holes are unstable, aligning with expectations from cosmic censorship and unitarity.

A substantial body of research has been dedicated to deriving the WGC from diverse theoretical frameworks. Within the context of string theory, significant progress has been made in understanding how the WGC emerges from the spectrum of charged states, modular invariance, and the swampland distance conjecture~\cite{Arkani-Hamed:2006emk,2401.14449,1903.06239,2201.08380,2409.10003,Bastian:2020egp}. Another fruitful avenue of investigation has focused on scattering amplitudes, where the WGC is linked to constraints on the low-energy behavior of gauge and gravitational interactions. Analytic properties of scattering amplitudes, such as positivity bounds and unitarity, have been shown to impose stringent conditions that align with the WGC, particularly in the context of effective field theories~\cite{Bellazzini:2019xts,Hamada:2018dde,Arkani-Hamed:2021ajd,Henriksson:2021ymi,Henriksson:2022oeu,Alberte:2020bdz,Tokareva:2025rta,Barbosa:2025smt}. These amplitude-based approaches provide a complementary perspective, rooted in the principles of quantum field theory and the consistency of S-matrix elements. Beyond these frameworks, additional support for the WGC has been obtained from a variety of theoretical considerations, such as holographic arguments \cite{Nakayama:2015hga,Harlow:2015lma,Benjamin:2016fhe,Montero:2016tif}, properties of the dimensional reduction~\cite{Brown:2015iha,Brown:2015lia,Heidenreich:2015nta,Heidenreich:2016aqi,Lee:2018urn}, and infrared consistency conditions, such as the absence of the global symmetries ~\cite{Cheung:2014ega,Andriolo:2018lvp,Bittar:2024xuc}.

From the perspective of black hole physics, the WGC has been explored through the lens of black hole thermodynamics \cite{Cheung:2018cwt}, extremality~\cite{Cottrell:2016bty,Hebecker:2017uix,Ma:2021opb,Abe:2023anf,DeLuca:2022tkm,Barbosa:2025uau}, and stability. The conjecture implies the requirement that extremal black holes must be able to decay, thereby avoiding remnants and preserving unitarity. Recent work has also examined the role of higher-curvature corrections and higher-dimensional operators in modifying the charge-to-mass ratio of Reissner--Nordstr\"om (RN) black holes~\cite{Cao:2022ajt,Cao:2022iqh,Aalsma:2019ryi}.

The stability of black holes is a critical issue, as unstable configurations could lead to violations of cosmic censorship or the formation of naked singularities. In this context, quasinormal modes (QNMs) play a pivotal role. QNMs are the characteristic damped oscillations of a perturbed black hole. Furthermore, the QNMs spectrum of RN black holes is sensitive to the extremal limit, where the black hole's charge approaches its mass ($Q \to M$). In this regime, the QNMs' frequencies exhibit distinctive behavior, such as the appearance of slowly damped modes, which can signal the onset of instabilities or the breakdown of perturbation theory. Higher-dimension EFT operators deform the black hole solution~\cite{Kats:2006xp,Cheung:2016yqr,Wang:2022sbp}. These deformations, affecting the QNMs, were recently studied for the Kerr black holes \cite{Cardoso:2018ptl,Cano:2021myl,Cano:2023jbk,Cano:2023tmv,Cano:2024ezp,Cano:2024jkd,Maenaut:2024oci,Melville:2024zjq,Cano:2024wzo,Cano:2025mht}. Recent works \cite{Boyce:2025fpr,Miguel:2023rzp} study EFT corrections to QNMs of RN black holes, including their near extremal regime. The latter is specifically interesting because the effects of EFT operators can be significantly enhanced. Moreover, in the extremal limit, the EFT description breaks down near the horizon \cite{Horowitz:2023xyl,Horowitz:2024dch,Chen:2024sgx}. 

For a comprehensive review of quasinormal modes of black holes in string theory and brane scenarios, we refer the reader to \cite{Konoplya:2011qq}.

In our work, we focus on the near-extremal limit of the EFT-corrected RN black hole, studying the stability of this geometry with massless probe scalar field perturbations. We study the effects of all EFT operators with four derivatives that the field redefinitions cannot eliminate. We found that, different from the previous studies of gravitational and electromagnetic perturbations, an effective potential for the massless scalar probe is sensitive to the coupling $ h_4 F_{\mu\nu}F_{\rho\sigma}R^{\mu\nu\rho\sigma}$.

It is well known that the Teukolsky equation, which describes spin-$s$ perturbations in the background generated by black holes in the Kerr-Newmann (KN) family, can be mapped to a confluent Heun equation (CHE), i.e., an ordinary differential equation (ODE) with two regular (Fuchsian) singularities and one irregular singularity \cite{Bianchi:2021mft}. This type of equation appears in plenty of geometric constructions, including but not limited to topological stars, black holes in AdS, D-branes, and fuzzballs. The topological star (a non-supersymmetric, horizonless 5D Einstein-Maxwell solution) is a Schwarzschild mimicker in the CHE class \cite{Bah:2020pdz,Bianchi:2023sfs,Heidmann:2023ojf,Bianchi:2025uis}, matching its large-distance physics but allowing possible small deviations \cite{Bianchi:2024vmi,Bianchi:2024rod,DiRusso:2025lip,Melis:2025iaw}. The Heun equation (HE) also governs AdS$_4$-KN perturbations \cite{Aminov:2023jve} and the Mathieu equation (doubly reduced doubly confluent Heun equation) describes D-branes \cite{Bianchi:2021xpr, Gregori:2022xks,Fioravanti:2021dce}. Some fuzzball geometries (e.g., D1-D5, JMaRT) also map to the reduced confluent Heun equation (RCHE) \cite{Bianchi:2022qph, Bianchi:2023rlt}. 
For a more extensive, though not complete, dictionary between certain gravitational solutions and the Heun-type equations, see \cite{Bianchi:2022wku}.

Quite recently, a correspondence between the QNMs spectral problem and the quantum Seiberg-Witten (qSW) curves for $N=2$ super-Yang-Mills (SYM) theories with gauge group $SU(2)$ was found \cite{Aminov:2020yma}. 

For some recent applications of qSW techniques to the reconstruction of the waveform emitted during the inspiral of a binary system of two compact objects and their equivalence with more established and well-known techniques in the literature, such as the Mano-Suzuki-Takasugi method used to solve the homogeneous CHE, see \cite{Bianchi:2024vmi,Cipriani:2025ikx}.

In the most general case, the qSW curve of the SYM with 4 mass hypermultiplets is exactly an HE with 4 regular singularities. Moreover, exploiting the Alday-Gaiotto-Tachikawa (AGT) correspondence \cite{Alday:2009aq}, the wave functions for BH solutions were related to correlators in the two-dimensional Liouville conformal field theory (CFT) involving degenerate fields, providing new tools to study further interesting quantities such as Tidal Love numbers \cite{DiRusso:2024hmd}, amplification factors \cite{Cipriani:2024ygw}, absorption coefficients \cite{Bianchi:2022qph}, etc. In this chain of correspondences, the powerful instantonic computation techniques, together with localization, reduce the mathematical problem of QNMs to a quantization condition.

From the computational point of view, a nearly extremal regime leads to certain challenges in applying known analytical and numerical methods of computing QNMs. The extremal limit corresponds to a confluence in which two regular singularities merge to form a single irregular singularity. From the perspective of differential equations, this transition is highly non-trivial, as the analytic structure of the solutions changes drastically: local Frobenius expansions around regular singular points are no longer valid. From the numerical integration viewpoint, this situation is particularly delicate, since standard methods that rely on stable series expansions or well-separated singularities become less reliable. Small numerical errors can be amplified by the presence of Stokes phenomena and exponential sensitivity associated with irregular points. Even the well-known Leaver method, based on the solution of a continued fraction \cite{Leaver:1985ax,Leaver:1990zz}, which is widely regarded as one of the most effective and accurate techniques for computing quasinormal modes, encounters serious difficulties in the extremal limit.

%Quite recently, a correspondence between the QNMs spectral problem and the quantum Seiberg-Witten (qSW) curves \cite{aminovgrassi} for $N=2$ super-Yang-Mills (SYM) theories with gauge group $SU(2)$ was found. In the most general case, the qSW curve of the SYM with 4 mass hypermultiplets is exactly an HE with 4 regular singularities. Moreover, exploiting the Alday-Gaiotto-Tachikawa (AGT) correspondence \cite{AGT}, the wave functions for BH solutions were related to correlators in the two-dimensional Liouville conformal field theory (CFT) involving degenerate fields, providing new tools to study further interesting quantities, such as Tidal Love numbers \cite{TLNGDR}, amplification factors \cite{chargeTS}, absorption coefficients \cite{D1d5GDR}, etc. In this chain of correspondences, the powerful instatonic computation techniques together with localization, the mathematical problem of QNMs is reduced to a quantization condition.

In this paper, we develop a systematic method allowing us to approximate the wave equation corrected by EFT operators by the analytically tractable CHE. This approximation is valid in the near-extremal limit. Using the analytic expression for the QNMs following from the Heun equation, we find the EFT corrections to the frequencies of the slowly damped modes. Remarkably, the first-order correction in deviation from the extremality is proportional to the same combination of Wilson coefficients emerging from the WGC requirement that a black hole with $M=Q$ has no naked singularity \cite{Arkani-Hamed:2006emk}. A causality condition formulated in \cite{Melville:2024zjq} as the damping of QNMs should always be larger than in pure GR, requires the same combination of the EFT couplings to be positive. 

The paper is organized as follows. 
\begin{description}

\item[Sec.~\ref{sec:EM}:] We introduce the Einstein-Maxwell EFT, corrections to the RN solution, and modification of the extremality condition.

\item[Sec.~\ref{RNsec}:] We review the duality relation between the problem of finding QNMs in RN black hole and quantum Seiberg-Witten curve in $N=2$ SYM gauge theory. We analytically reproduce the frequencies of the Zero Damping Modes in the near extremal limit from the quantisation condition on the Seiberg-Witten cycle in the decoupling limit of SYM with three flavours.

\item[Sec.~\ref{sec:ZDMinEFT}:] We construct an approximation for the effective potential of scalar wave in EFT which is valid in the near extremal limit, and leads to a similar form of Confluent Heun Equation, compared to the one discussed in the previous Section. We obtain the EFT-corrected QNM frequencies analytically from matching the approximated potential to the corresponding quantum Seiberg-Witten curve.

\item[Sec.~\ref{sec:PRM}:] We compute the Lyapunov exponent and Prompt Ringdown Modes for scalar wave in RN geometry using WKB approximation, Leaver method, and numerical integration.

\item [Sec.~\ref{sec:conclusions}:]  We summarize our results and discuss their relevance for such consistency conditions on the EFT couplings, as causality requirements and WGC. 

\end{description}

\section{Einstein-Maxwell effective theory}
\label{sec:EM}
\subsection{Action and deformed solution for RN black hole.}

As perturbative quantum gravity based on the Einstein-Hilbert action is non-renormalisable in the framework of quantum field theory, physics below the Planck scale can be parametrized by the derivative expansion containing all the combinations of operators compatible with diffeomorphism invariance. If gravity is coupled to the electromagnetic field, the operators mixing the Riemann tensor with the electromagnetic field strength can also emerge as counterterms. Although there are a lot of covariant combinations and contractions that can be written, there is a very limited set of operators of the lowest dimension that remain after perturbative field redefinitions \cite{Ruhdorfer:2019qmk}. The action of the Einstein-Maxwell effective theory in four dimensions up to the terms containing four derivatives can be written as
\be
S_4=\int d^4x\sqrt{-g}\Big[\frac{R}{2\k}-\frac{1}{4}F_{\m\n}F^{\m\n}+g_4(F_{\m\n}F^{\m\n})^2+h_4 R^{\mu\nu\rho\sigma}F_{\mu\nu}F_{\rho\sigma}+f_4 (F_{\mu\nu}\tilde{F}^{\mu\nu})^2\Big]
\ee
where the field strength is $F_{\m\n}=\nabla_\m A_\n-\nabla_\n A_\m$ and $\nabla_\m $ is the covariant derivative, $\tilde{F}_{\mu\nu}=\epsilon_{\mu\nu\rho\sigma}F^{\mu\nu}$ and $\epsilon_{0,1,2,3}=1$. The other terms of the same dimension can be reduced to the total derivatives or removed by the perturbative field redefinitions \cite{Ruhdorfer:2019qmk,Basile:2024oms}. The latter transforms the redundant operators into the ones suppressed by higher powers of the EFT breakdown scale. This expansion is valid only for the energy scales or curvatures of the spacetime not exceeding the EFT cutoff scale, which can be roughly estimated as the minimal scale among $g_4^{-1/4},~h_4^{-1/4},~f_4^{-1/4}$.

The field equations following from the Einstein-Maxwell EFT truncated at four-derivative operators are
\begin{align}
\label{EFEg}
R_{\m\n}&-\frac{1}{2}g_{\m\n}R=\k\Big[{F_\m}^\a F_{\n\a}-\frac{1}{4} F^2 g_{\m\n} +g_4F^2\left(F^2 g_{\m\n}-8{F_{\m}}^\a F_{\nu\a}\right)\nonumber\\
&+8f_4\left(\tilde{F}_{\mu\rho}{F_\nu}^\rho+\tilde{F}_{\nu\rho}{F_\mu}^\rho\right) F_{\alpha\beta}\tilde{F}^{\alpha\beta}\nonumber\\
&+\kappa h_4\left(g_{\mu\nu}F_{\alpha\beta}F_{\gamma\delta}R^{\alpha\beta\gamma\delta}-6F_{\alpha\nu}F^{\gamma\delta}{{R^\alpha}_{\mu\gamma\delta}}-4\nabla_\beta \nabla_\alpha {F^\alpha}_\mu{F^\beta}_\nu\right)\Big],
\end{align}
\begin{align}\label{EFEf}
\nabla_\m F^{\m\n}&=8g_4(  F^2\nabla_\m F^{\m\n}+F^{\m\n}\nabla_\m F^2)+8 f_4\left(F_{\alpha\beta}\tilde{F}^{\alpha\beta}\nabla_\rho \tilde{F}^{\nu\rho}+2\tilde{F}^{\nu\rho}\tilde{F}_{\alpha\beta}\nabla_\rho F^{\alpha\beta}\right)\nonumber\\
&+4h_4 \nabla_\nu(R^{\alpha\beta\mu\nu}F_{\alpha\beta}).
\end{align}

These equations can be solved perturbatively in EFT coefficients as a deformation of the very well-known Reissner-Nordstr\"om (RN) solution in $d=4$. With this purpose in mind, the ansatz for a spherically symmetric solution can be constructed as follows\footnote{As it is shown in \cite{Kats:2006xp,Campanelli:1994sj}, for the deformed RN solution, the functions in front of $dt^2$ and $dr^2$ remain inverse to each other.}
\be\label{solpert}
ds^2=-G(r)dt^2+\frac{dr^2}{G(r)}+r^2(d\q^2+\sin^2\q d\phi^2), 
\ee
\be \label{EFTsol1} G(r)=1-\frac{2M}{r}+\frac{\k Q^2}{2r^2}+\sum_{n=f_4,g_4,h_4}n\, \mathcal{S}_n(r),
\ee
\be\label{EFTsol2}
F_{\m\n}=\left(\frac{Q}{r^2}+\sum_{n=f_4,g_4,h_4}\, n\,\delta A_n(r)\right)dt\wedge dr.
\ee 
Working in Planck units, where $\kappa = 2$, which will be employed throughout the paper, the perturbative solution at first order in $g_4$ compatible with asymptotic flatness was obtained in \cite{Kats:2006xp,Campanelli:1994sj}.
\be
\delta A_{g_4}(r)=-\frac{16Q^3}{r^6}\quad,\quad \mathcal{S}_{g_4}(r)=-\frac{8Q^4}{5r^6},
\ee
\be
\delta A_{h_4}(r)=\frac{16 M Q}{r^5}-\frac{24Q^3}{r^6}\quad,\quad \mathcal{S}_{h_4}(r)=-\frac{64Q^4}{5r^6}+\frac{28 M Q^2}{r^5}-\frac{16Q^2}{r^4},
\ee
\be
\delta A_{f_4}(r)=0\quad,\quad \mathcal{S}_{f_4}(r)=0.
\ee
We see that only $g_4$ and $h_4$ terms contribute as corrections to the RN solution.

\subsection{Extremality condition and WGC}

The extremal regime verifies when the function $G(r)$ fully characterizing the metric shows a double zero. The six roots of $G(r)$ are such that two of them are real and positive, and the largest correspond to the real black hole horizon. They approach RN horizons $r_\pm=M\pm\sqrt{M^2-Q^2}$ in the limit of $g_4,~h_4\rightarrow 0$. In small $g_4$ and $h_4$ expansion the horizons read
\begin{align}
\label{R1+-}
R_+^{(1)}&=r_++\frac{4g_4Q^4}{4r_+^4\sqrt{M^2-Q^2}}+\frac{2h_4Q^2(5M r_+-4Q^2)}{5r_+^4\sqrt{M^2-Q^2}}+O\left(\frac{g_{EFT}^2}{(M^2-Q^2)^{3/2}},\right)\,,\nonumber\\
R_-^{(1)}&=r_--\frac{4g_4Q^4}{5r_-^4\sqrt{M^2-Q^2}}+\frac{2h_4Q^2(4Q^2-5M r_-)}{5r_-^4\sqrt{M^2-Q^2}}+O\left(\frac{g_{EFT}^2}{(M^2-Q^2)^{3/2}},\right)\,.
\end{align}

Here $g_{EFT}^2$ stands for a combination quadratic in $g_4,~h_4$. Remarkably, the EFT correction to the location of the horizon is expressed as a series expansion in parameters 
\be
\frac{g_4}{M^2-Q^2},~\frac{h_4}{M^2-Q^2}.
\ee
In the extremal regime, the series expansion \eqref{R1+-} diverges, which means that for given values of EFT couplings, it cannot be used for the black hole parameters too close to the extremality condition. In addition, in this regime, non-linear effects are becoming important, leading to the breakdown of our perturbative expansion in EFT couplings \cite{Horowitz:2022mly,Horowitz:2023xyl,Horowitz:2024dch,Horowitz:2024kcx,Chen:2024sgx,DelPorro:2025fiu}. In this paper, we mainly focus on the perturbative regime
\be
\frac{g_4}{M^2-Q^2}\ll 1,~\frac{h_4}{M^2-Q^2}\ll 1.
\ee

The extremality condition $Q=M$ receives corrections from the EFT operators. Indeed,
equating $R_+^{(1)}=R_-^{(1)}$ and expanding in the couplings $g_4$ and $h_4$ at linear order, we obtain a condition\footnote{Let us stress here that this condition was obtained perturbatively in the EFT couplings.} 
\be
M=Q-\frac{2g_4+h_4}{5Q}\,.
\ee
This modification plays a major role in the formulation of the WGC proposed in \cite{ArkaniHamed:2006mb}. Namely, as follows from this conjecture, a black hole with $M=Q$ should still be a solution without a naked singularity. It is possible only if
\be\label{WGCcond}
2g_4+h_4\ge0\,.
\ee
This condition provides one of the important EFT consistency constraints following from black hole physics.

\subsection{Scalar wave equation}

In this paper, we examine perturbations of a massless scalar field on top of the EFT-corrected RN black hole geometry. The wave equation for a massless scalar in the background \eqref{solpert} is of the form
\be
\Box \Phi(t,r,\q,\phi)=\frac{1}{\sqrt{-g}}\partial_\mu\Big[\sqrt{-g} g^{\m\n}\partial_\n \Big]\Phi(t,r,\q,\phi)=0\,.
\ee
The ansatz dictated by the spherical symmetry
\be
\Phi(t,r,\q,\phi)={\rm exp}\left(-{\rm i}\omega t+{\rm i} m\phi\right)R(r)S(\theta)
\ee
allows us to separate the dynamics. The angular equation 
\be
\Big[\frac{1}{\sin(\q)}\partial_\q\left(\sin\q \partial_\q\right)-\frac{m^2}{\sin^2\q}\Big]S(\q)=-\ell(\ell+1)S(\q)
\ee
can be solved in terms of Legendre polynomials. The radial equation becomes
\be
r^2 G(r)R''(r)+r\left(2G(r)+rG'(r)\right)R'(r)+\left(\frac{r^2\omega^2}{G(r)}-\ell(\ell+1)\right)R(r)=0\,,
\ee
which can be rewritten in a canonical (or normal) form 
$$   R(r)=\frac{\psi(r)}{r \sqrt{G(r)}},\qquad \psi''(r)+Q_W(r,\omega)\psi(r)=0\,,$$
\be\label{scalwaveeq}
Q_W(r,\omega)=\frac{r^2(4\omega^2+G'(r)^2)-2G(r)(2\ell(\ell+1)+2r G'(r)+r^2 G''(r))}{4r^2G(r)^2}\,.
\ee

For the EFT-corrected solution \eqref{EFTsol1}, \eqref{EFTsol2} we have
\begin{equation}\label{QWnum}
\begin{split}
& Q_W^{{\rm num}}=\frac{8 g_4 Q^4 \left(5 r^4 \left(\left(\ell^2{+}\ell{+}15\right) r^2{-}24 M r{+}10 Q^2\right){-}8 h_4 Q^2 \left({-}175 M r{+}96 Q^2{+}90 r^2\right)\right)}{4 r^2 \left({-}\frac{8 g_4 Q^4}{5 r^6}{+}h_4 \left(\frac{28 M Q^2}{r^5}{-}\frac{64 Q^4}{5 r^6}{-}\frac{16 Q^2}{r^4}\right){-}\frac{2 M}{r}{+}\frac{Q^2}{r^2}{+}1\right)^2} \\
& + \frac{20 h_4 Q^2 r^4 \left(5 r^2 (7 M{-}4 r) \left(15 M{-}\left(\ell^2{+}\ell{+}6\right) r\right){+}2 Q^2 r (2 (4 \ell (\ell{+}1){+}75) r{-}297 M){+}160 Q^4\right)}{4 r^2 \left({-}\frac{8 g_4 Q^4}{5 r^6}{+}h_4 \left(\frac{28 M Q^2}{r^5}{-}\frac{64 Q^4}{5 r^6}{-}\frac{16 Q^2}{r^4}\right){-}\frac{2 M}{r}{+}\frac{Q^2}{r^2}{+}1\right)^2} \\
& + \frac{25 r^{10} \left(-\left(\ell^2+\ell+1\right) Q^2+2 \ell (\ell+1) M r-\ell (\ell+1) r^2+M^2+r^4 \omega ^2\right)}{4 r^2 \left(-\frac{8 g_4 Q^4}{5 r^6}+h_4 \left(\frac{28 M Q^2}{r^5}-\frac{64 Q^4}{5 r^6}-\frac{16 Q^2}{r^4}\right)-\frac{2 M}{r}+\frac{Q^2}{r^2}+1\right)^2}.
\end{split}
\end{equation}
Unlike the original RN potential for the scalar wave, this equation, in general, can be solved numerically and in the WKB approximation. However, we will show that in the near extremal limit, this equation can be well approximated by the confluent Heun-type equation, which allows for an analytical expression of EFT corrections to the QNMs.

\section{Zero damping modes in Reissner-Nordstr\"om geometry and quantum Seiberg-Witten curves}\label{RNsec}
The exact Reissner-Nordstr\"om solution can be recovered by taking $g_4=0,~h_4=0$ in \eqref{solpert}. The scalar massless wave equation appears to be
\be\label{scawaeq}
(r{-}r_+)(r{-}r_-)R''(r){+}(2r{-}r_+{-}r_-)R'(r){+}\Big[\frac{r^4\omega^2}{(r{-}r_+)(r{-}r_-)}{-}\ell(\ell{+}1)\Big]R(r){=}0,
\ee
which can be rewritten in a canonical form by the following redefinition of the wave function
\be
R(r)=\frac{\psi(r)}{\sqrt{(r-r_+)(r-r_-)}}.
\ee
so that the transformed ODE becomes 
$$\psi''(r)+Q_{W,RN}(r,\omega)\psi(r)=0,$$
where the effective potential is
\be\label{RNcan}
Q_{W,RN}(r,\omega)=\frac{r^4\omega^2-\ell(\ell+1)(r-r_+)(r-r_-)+\frac{1}{4}(r_+-r_-)^2}{(r-r_+)^2(r-r_-)^2}\,.
\ee
It is known that all perturbations of the geometries of black holes in the Kerr-Newman family admit wave equations that can be mapped to Heun equations. In particular, all black holes in the Kerr-Newman family can be mapped to confluent Heun equations (CHE). As we will show in the next subsection, Heun equations also arise from the Seiberg-Witten (SW) curve embedded in the Nekrasov-Shatashvili background. Ultimately, in suitable coordinates, it will be possible to establish a precise dictionary between the SW curve parameters (five in the case of the CHE) and the differential equation obtained from black-hole perturbations.
%{\color{red}[Here the transition to SW curve should be explained in 4-5 sentences,if possible, SYM should be defined]}
\subsection{Quantum Seiberg-Witten curves for $\mathcal{N}=2$ SYM with flavours}
In this Section, we will very briefly derive the quantum version of the SW curve starting from the \lq \lq classical" case for the theory $SU(2)$ $\mathcal{N}=2$ supersymmetric Yang-Mills (SYM) with $N_f=4$. Then we will derive the theory with three flavors by taking the so-called decoupling limit, obtaining the confluent Heun equation describing the scalar wave equation \eqref{RNcan} in RN geometry. For a more detailed explanation of the topic, we refer to \cite{Bianchi:2021mft,Bianchi:2021xpr}.

The field content of $\mathcal{N}=2$ $SU(2)$ SYM consists of a gauge boson, two gauginos, and a complex scalar field in the adjoint representation. For our purposes, we also include additional massive hypermultiplets in the fundamental representation. 

The main feature of this supersymmetric theory consists of the exact knowledge of the non-perturbative low-energy effective dynamics of the theory, which in particular contains the effective renormalized gauge coupling $g_{eff}$ and theta-angle, $\theta_{eff}$:
\be
\tau=\frac{\theta_{eff}}{\pi}+\frac{8\pi i}{g_{eff}^2}=\frac{8\pi i}{g_0^2}+\frac{2i}{\pi}\log\left(\frac{a^2}{\Lambda^2}\right)-\frac{i}{\pi}\sum_{i=1}^\infty c_i\left(\frac{\Lambda}{a}\right)^{4i}
\ee
where $\Lambda$ is the dynamically generated scale at which the gauge coupling becomes strong and $a$ is the Higgs field \cite{Lerche:1996xu}. 

On the other side, it is known that the low-energy effective dynamics of supersymmetric gauge theories are described in terms of a function called the prepotential. As a result, for the $N=2$ $SU(2)$ SYM theory the prepotential $\mathcal{F}$ is known exactly and is a holomorphic function.

The vacuum expectation value of the adjoint scalar breaks the gauge symmetry from $SU(2)$ to $U(1)$, and the resulting quantum dynamics is encoded in the Seiberg-Witten curve \cite{Witten:1997sc,Seiberg1994ElectricM}. In the presence of four hypermultiplets with masses $m_i$, the curve takes the form
\be\label{classSW}
q y^2 P_L(x)+y P_0(x)+P_R(x)=0,
\ee
where
\be\label{P0RL}
P_0(x)=x^2-u+q p_0(x),\quad P_R(x)=(x-m_1)(x-m_2),\quad P_L(x)=(x-m_3)(x-m_4).
\ee
Here $q=e^{2\pi i \tau}$ is the gauge parameter and $u$ is the Coulomb branch modulus, and $p_0(x)$ is a quadratic function in $x$ determined below \eqref{p0}. 

Solutions of \eqref{classSW} for $y$ have the form
\be
y_\pm=\frac{1}{2qP_L(x)}\left(-P_0\pm\sqrt{P_0^2-4qP_LP_R}\right)\,.
\ee
Thus, \eqref{classSW} is an elliptic curve and can be viewed as a double cover of the complex plane with four branch points defined by
\be
P_0^2-4q P_LP_R=\prod_{i=1}^4(x-e_i)\,.
\ee
The periods of the elliptic curve are defined as,
\be
a=\oint_\alpha \lambda_0\quad,\quad a_D=\oint_\beta\lambda_0
\ee
where $\alpha$ and $\beta$ being the two fundamental cycles, and $\lambda_0$ is the the SW differential defined as
\be
\lambda_0=\frac{1}{2}(\lambda_+-\lambda_-)\quad,\quad \lambda_\pm=\frac{1}{2\pi i}x\partial_x \ln y_\pm(x) dx\,.
\ee
The quantum version of this setup consists of promoting the variables appearing in the SW curve \eqref{classSW} to be the operators satisfying the commutation relation 
\be\label{NScomm}
[\hat{x},\ln \hat{y}]=\hbar\,.
\ee
This is called the Nekrasov-Shatashvili (NS) background \cite{Nekrasov:2009rc} corresponding to non-commutative space\footnote{In this construction $\hbar$ has nothing to do with the Planck constant. It represents simply the deformation parameter of the background.}. Starting from \eqref{classSW}, the quantum curve can be written as follows
\be\label{swcurveA}
\Big[q\hat{y}^{\frac{1}{2}}P_L(\hat{x})\hat{y}^{\frac{1}{2}}+P_0(\hat{x})+\hat{y}^{-\frac{1}{2}}P_R(\hat{x})\hat{y}^{-\frac{1}{2}}\Big]U(x)=0\footnote{U is a function introduced to allow the differential operators to act.},
\ee
 where $P_0$, $P_L$, $P_R$ are given exactly as in \eqref{P0RL} and
\be\label{p0}
p_0(x)=x^2-\left(x+\frac{\hbar}{2}\right)\sum_i m_i+u+\sum_{i<j}m_i m_j+\frac{\hbar^2}{2}
\ee
With the use of \eqref{NScomm}, the \lq\lq classical" curve \eqref{classSW} becomes an ordinary differential equation in $y$ variable
\be\label{quantSW}
\Big[qy^2P_L\left(\hat{x}{+}\frac{\hbar}{2}\right){+}y P_0(\hat{x}){+}P_R\left(\hat{x}{-}\frac{\hbar}{2}\right)\Big]U(y){=}\Big[A(y)\hat{x}^2{+}B(y)\hat{x}{+}C(y)\Big]U(y){=}0
\ee
with
\begin{eqnarray}
&&A=(1+y)(1+q\,y)\quad,\quad B=-m_1-m_2-\hbar+q y\Big[y(\hbar-m_3-m_4)-\sum_i m_i\Big]\\
&&C{=}\left(m_1{+}\frac{\hbar}{2}\right)\left(m_2{+}\frac{\hbar}{2}\right){-}u y{+}q y\Big[u{+}\sum_{i<j}m_im_j{-}\frac{\hbar}{2}\sum_i m_i{+}\frac{\hbar^2}{2}{+}y\left(m_3{-}\frac{\hbar}{2}\right)\left(m_4{-}\frac{\hbar}{2}\right)\Big]\,.\nonumber
\end{eqnarray}
The differential equation \eqref{quantSW} can be brought to the normal (or canonical) form with the rescaling
\be
U(y)=\frac{1}{\sqrt{y}}e^{-\frac{1}{2\hbar}\int^y\frac{B(y')}{y'A(y')}dy'}\Psi(y)\,.
\ee
The potential becomes
\begin{eqnarray}\label{22curve}
&&\Psi''(y)+Q_{SW}(y) \Psi(y)=0\\
&&Q_{SW}(y)=\frac{4C A-B^2+2\hbar y(BA'-AB')+\hbar^2A^2}{4\hbar^2u^2A^2}\,.\nonumber
\end{eqnarray}
More explicitly, \eqref{22curve} is
\begin{eqnarray}
&&Q_{2,2}(y)=\frac{\hbar^2-(m_1-m_2)^2}{4\hbar^2y^2}+\frac{\hbar^2-(m_1+m_2)^2}{4\hbar^2(1+y)^2}+\frac{q^2(\hbar^2-(m_3+m_4)^2)}{4\hbar^2(1+q y)^2}\nonumber\\
&+&\frac{1}{4\hbar^2y(1+y)(1+q y)}\Big[q \left(2 m_1^2 y+2 m_2^2 y+4 m_3 m_4 y-2\left( m_1+ m_2+ m_3+ m_4\right)
   \hbar\right. \nonumber\\
   &{+}&\left.2 m_1 m_3{+}2 m_2 m_3{+}2 m_1 m_4{+}2 m_2 m_4{+}4 m_3 m_4{+}4 u{+}(1{-}2 y) \hbar
   ^2\right){+}2 m_1^2{+}2 m_2^2{-}4 u{-}\hbar ^2\Big]\,.\nonumber\\
\end{eqnarray}
The quantum curve with $N_f=3$ can be obtained in the so-called decoupling limit
\be
q\to 0,\quad m_4\to \infty,\quad q'=-qm_4={\rm const},
\ee
so that we obtain
\begin{equation}\label{Q21def}
\begin{split}
Q_{1,2}(y) &= -\frac{q'^2}{4\hbar^2} + \frac{\hbar^2 - (m_1 - m_2)^2}{4\hbar^2 y^2} + \frac{\hbar^2 - (m_1 + m_2)^2}{4\hbar^2 (1 + y)^2} \\
&\quad - \frac{m_3 q'}{\hbar^2 y} + \frac{2(m_1^2 + m_2^2) - \hbar^2 - 4u + 2q'(\hbar - m_1 - m_2)}{4\hbar^2 y(1 + y)}\,.
\end{split}
\end{equation}

The differential equation
\be\label{qSWcurve3}
\Psi''(y)+Q_{1,2}(y) \Psi(y)=0
\ee
with the potential \eqref{Q21def}, can be reduced to the form of the Confluent Heun equation, which admits an analytic solution. This equation resembles the form of \eqref{RNcan}, which means that the problem of finding QNMs for the RN geometry can be mapped to the properties of the quantum SW curve. Indeed,
%\subsection{Reissner-Nordstr\"om dictionary}
after the change of variables in \eqref{RNcan}
\be
y=\frac{r-r_+}{r_+-r_-}
\ee
we obtain the dictionary with the qSW curve \eqref{Q21def} for $N_f=3$,
$$
q=2i(r_+-r_-)\omega,\quad m_1=\frac{i(r_+^2+r_-^2)\omega}{r_+-r_-},\quad m_2=-m_3=-i(r_++r_-)\omega
$$
\be\label{dict21}
u=\left(\ell+\frac{1}{2}\right)^2-\omega(i(r_--r_+)+(r_-+r_+)^2\omega)\,.
\ee
From now on, since there is no possibility of confusion, we will denote the gauge coupling parameter of the $N_f=3$ theory simply by $q$, dropping the prime symbol appearing in \eqref{Q21def}. We will also set $\hbar=1$, as the correct dependence on the deformation parameter $\hbar$ can be recovered by dimensional analysis if needed.

\subsection{Quantum Seiberg-Witten cycles and QNM frequencies}
Using the commutation relation \eqref{NScomm} and setting $\hat{y}=e^{-\hbar \partial_x}$, we can recast the qSW curve \eqref{quantSW} in the following form 
\be\label{qSW2}
\Big[qP_L\left(x-\frac{1}{2}\right)\hat{y}+P_0(x)+P_R\left(x+\frac{1}{2}\right)\hat{y}^{-1}\Big]\tilde{U}(x)=0\,.
\ee
Introducing the functions
\be
W(x)=\frac{1}{P_R\left(x+\frac{1}{2}\right)}\frac{\tilde{U}(x)}{\tilde{U}(x+1)}\quad,\quad M(x)=q P_L\left(x-\frac{1}{2}\right)P_R\left(x-\frac{1}{2}\right)\quad\,,
\ee
Eq. \eqref{qSW2} can be written in the form
\be
qM(x)W(x)W(x-1)+P_0(x)W(x)+1=0\,.
\ee
The previous difference equation can be solved in two ways: first in $W(x)$, and then in $W(x-1)$. By shifting the latter expression and comparing the two solutions, we are finally left with the following continued fraction
\be\label{diffeqSW}
P_0(a)=\frac{M(a+1)}{P_0(a+1)-\frac{M(a+2)}{P_0(a+2)-\dots}}+\frac{M(a)}{P_0(a-1)-\frac{M(a-1)}{P_0(a-1)-\dots}},
\ee

Equation \eqref{diffeqSW} can be solved for $u(a,q)$ perturbatively in the gauge coupling $q$, getting
\begin{align}
\label{uvsa}
&u(a,q) =  a^2 + q \left[\frac{1}{2}(1-m_1-m_2-m_3) - \frac{2 m_1 m_2 m_3}{4 a^2 - 1}\right] + \frac{q^2}{128 (a^2-1)} \left[4 a^2 -5 + \notag \right.\\
&\left.+ 4(m^2_1+m^2_2+m^2_3) - \frac{48 (m_1^2 m_2^2 + m_2^2 m_3^2 + m_1^2 m_3^2)}{4a^2 -1} + \frac{64 m^2_1 m^2_2 m^2_3(20 a^2+7)}{(4 a^2 -1)^3} \right]+\dots
\end{align}
We can invert this series in order to obtain the qSW cycle
\begin{align}
\label{avsu}
&a(u,q) = \sqrt{u} + \frac{q}{4 \sqrt{u}} \left(\frac{4 m_1 m_2 m_3}{4u-1} + m_1 + m_2 + m_3 - 1\right) + \\\nn
&-\frac{q^2}{256 \sqrt{u}} \left(\frac{1024 m_1^2 m_2^2 m_3^2}{(4
   u-1)^3}-\frac{256 m_1 m_2 \left(m_1 \left(m_2 m_3-2\right)-2
   \left(m_2+m_3-1\right)\right) m_3}{(1-4 u)^2}+\right.\\\nn
   &\left.+\frac{8
   \left(m_2+m_3+m_1 \left(1-4 m_2
   m_3\right)-1\right){}^2}{u}+\frac{\left(4 m_1^2-1\right)
   \left(4 m_2^2-1\right) \left(4 m_3^2-1\right)}{u-1}+\right.\\\nn
   &\left.+\frac{64
   \left(\left(\left(1{-}12 m_3^2\right) m_2^2{+}4 m_3
   m_2{+}m_3^2\right) m_1^2{+}4 m_2 m_3 \left(m_2{+}m_3{-}1\right)
   m_1{+}m_2^2 m_3^2\right)}{4 u{-}1}{+}4\right)+\dots
\end{align}
The Nekrasov-Shatashvili (NS) prepotential $\mathcal{F}_{NS}(a,q)$ for N=2 SYM theory can be obtained by integrating the quantum Matone relation \cite{Matone:1995rx,Flume:2004rp}
\be\label{matonerel}
u=-q \frac{\partial \mathcal{F}_{NS}(a,q)}{\partial q}.
\ee
Furthermore, we have to include a $q$-independent term representing the one-loop correction to the prepotential, which is obtained by integrating out the heavy fields around the vacuum. As a result, in $\mathcal{N}=2$ SYM, due to supersymmetry, contributions from higher-order loops vanish. Therefore, the prepotential has the following structure:
\be\label{FNS}
\mathcal{F}_{NS}(a,q)=\mathcal{F}_{tree}(a,q)+\mathcal{F}_{1-loop}(a)+\mathcal{F}_{inst}(a,q),
\ee
where
\begin{align}
\label{Prepotential}
& \mathcal{F}_{\text{tree}}(a,q) = -a^2 \log q, \notag \\
& \mathcal{F}_{\text{inst}}(a,q) = q \frac{1{-}m_2{-}m_3{+}4 a^2 (m_1 {+} m_2 {+} m_3 {-}1) {+} m_1 (4 m_2 m_3 {-} 1)}{2(4a^2{-}1)} {+} \dots,\\\nn
&\frac{\partial \mathcal{F}_{\text{one-loop}}(a)}{\partial a} = \log \left[\frac{\Gamma^2(1+2a)}{\Gamma^2(1-2a)} \prod_{i=1}^3 \frac{\Gamma\left(\frac{1}{2}+m_i-a\right)}{\Gamma\left(\frac{1}{2}+m_i+a\right)}\right].
\end{align}
As it usually happens in Euclidean Yang-Mills (YM) theory, instantons are classical and non-perturbative solutions of YM equations, and they encode the effects of topologically nontrivial gauge configurations, providing information about strong-coupling dynamics and the exact low-energy behavior of the theory.

Starting from the prepotential, the $a_D-$period can be defined as
\be\label{addef}
a_D=-\frac{1}{2\pi i} \frac{\partial \mathcal{F}_{NS}}{\partial a}\,.
\ee
This period is subject to the quantization condition
\be\label{condquantqnm}
a_D=n\quad,\quad n=0,1,2,...
\ee
This condition can be mapped to the problem of finding QNMs in the effective potential described by the same confluent Heun equation. In the references \cite{Bonelli:2022ten,Consoli:2022eey}, closed-form expressions for the connection formulae of Heun equations were derived. By imposing boundary conditions compatible with QNMs, the in-going solution at the horizon can be analytically continued to infinity, where it decomposes into two contributions corresponding to in-going and out-going waves. Requiring the coefficient of the in-going wave at infinity to vanish leads to the same quantization condition as \eqref{condquantqnm}  involving the dual cycle
\be\label{condquantqnm1}
a_D=n\quad,\quad n=0,1,2,...
\ee
The latter quantization condition for QNMs can be solved numerically for complex $\omega$'s, providing a value for the QNM frequencies.

\subsection{Matching solutions in decoupling limit $q\rightarrow 0$}\label{MAE}
Let us consider our differential equation describing scalar waves in RN in terms of the qSW curve appearing in \eqref{qSWcurve3} with the effective potential \eqref{Q21def}. The near-horizon regime is described by the limit of vanishing coupling $q=0$. In this regime, the CHE described by the $N_f=3$ qSW curve becomes a hypergeometric equation whose solutions are
\begin{eqnarray}
\psi_H(y){=}&&A_1y^{\frac{1}{2}(1{+}m_1{-}m_2)}(1{+}y)^{\frac{1}{2}(1{+}m_1{+}m_2)}{}_2F_1\Big[\frac{1}{2}{+}m_1{-}\sqrt{u},\frac{1}{2}{+}m_1{+}\sqrt{u},1{+}m_1{-}m_2,{-}y\Big]\nonumber\\
&+&(1\longleftrightarrow2)\,.
\end{eqnarray}
According to the dictionary \eqref{dict21}, the solution with ingoing boundary conditions at the horizon $y=0$ is
\be
\psi_{H,in}(y){=}y^{\frac{1}{2}(1{-}m_1{+}m_2)}(1{+}y)^{\frac{1}{2}(1{+}m_1{+}m_2)}{}_2F_1\Big[\frac{1}{2}{+}m_2{-}\sqrt{u},\frac{1}{2}{+}m_2{+}\sqrt{u},1{-}m_1{+}m_2,{-}y\Big]\,.
\ee
The latter equation can be analytically continued at infinity,
\be\label{match1}
\psi_{H,in}(y)\underset{y\rightarrow\infty }{\sim}y^{\frac{1}{2}-\sqrt{u}}+y^{\frac{1}{2}+\sqrt{u}}\frac{\Gamma(\frac{1}{2}-m_1-\sqrt{u})\Gamma(\frac{1}{2}+m_2-\sqrt{u})\Gamma(2\sqrt{u})}{\Gamma(\frac{1}{2}-m_1+\sqrt{u})\Gamma(\frac{1}{2}+m_2+\sqrt{u})\Gamma(-2\sqrt{u})}.
\ee
In the far zone region, the potential can be approximated as follows,
\be
Q_{1,2}\underset{y\rightarrow\infty}{\sim}-\frac{q^2}{4}-\frac{m_3 q}{y}+\frac{1-4u}{4y^2}\,.
\ee
The corresponding solution can be written in terms of confluent hypergeometric functions,
\be
\psi_{\infty}(y)=\sum_{\alpha=\pm}A_\alpha e^{\alpha\frac{q y}{2}}(q y)^{\frac{1}{2}+\sqrt{u}}U\Big[\frac{1}{2}-\alpha m_3+\sqrt{u},1+2\sqrt{u},-\alpha y\Big]\,.
\ee
The solution with outgoing behavior at infinity is
\be
\psi_{\infty,out}(y)=e^{\frac{q y}{2}}(q y)^{\frac{1}{2}+\sqrt{u}}U\Big[\frac{1}{2}-m_3+\sqrt{u},1+2\sqrt{u},- y\Big]\,.
\ee
The previous solution can be analytically continued at the horizon,
\be\label{match2}
\psi_{\infty,out}(y)\underset{y\rightarrow 0}{\sim }y^{\frac{1}{2}-\sqrt{u}}+y^{\frac{1}{2}+\sqrt{u}}\frac{(-q)^{2\sqrt{u}}\Gamma(\frac{1}{2}-m_3+\sqrt{u})\Gamma(-2\sqrt{u})}{\Gamma(\frac{1}{2}-m_3-\sqrt{u})\Gamma(2\sqrt{u})}\,.
\ee
Comparing the asymptotic behaviors, we obtain the following matching condition
\be\label{matching}
({-}q)^{2\sqrt{u}}\frac{\Gamma({-}2\sqrt{u})^2\Gamma(\frac{1}{2}{-}m_1{+}\sqrt{u})\Gamma(\frac{1}{2}{+}m_2{+}\sqrt{u})\Gamma(\frac{1}{2}{-}m_3{+}\sqrt{u})}{\Gamma(2\sqrt{u})^2\Gamma(\frac{1}{2}{-}m_1{-}\sqrt{u})\Gamma(\frac{1}{2}{+}m_2{-}\sqrt{u})\Gamma(\frac{1}{2}{-}m_3{-}\sqrt{u})}{=}1\,.
\ee
This last expression, which is also present in \cite{Yang:2012pj,Yang:2013uba}, is exactly the quantization of $a_D$ appearing in \eqref{FNS},\eqref{Prepotential}, and \eqref{addef} at the lowest order in $q$ 

\subsection{Analytic expression for slowly damped modes in near extremal limit}

In the limit of weak coupling $q\to 0$, in order to satisfy the matching condition \eqref{matching}, the decrease of $q^{2\sqrt{u}}$ must be compensated by the divergence of a Gamma function in the numerator. It turns out that the correct choice is to require
\be\label{condmatch}
\frac{1}{2}-m_1+\sqrt{u}=-n,
\ee
where $n$ can be interpreted as the overtone number. One can see from the dictionary \eqref{dict21} that $q$ is small in the near extremal limit. Setting $r_-=r_+-\delta$ and expanding for small $\delta>0$, we obtain from \eqref{condmatch},
\be\label{omegamatch}
\omega_{match}\sim-i\frac{\ell+1+n}{2r_+^2}(r_+-r_-). 
\ee
Thus, we obtained that in the near extremal limit there is a branch of quasinormal modes with the parametrically small (i. e., proportional to $(r_+-r_-)$) imaginary parts. These modes are often referred to as zero-damped modes (ZDM), which are widely discussed in the works \cite{Brito:2015oca,Bianchi:2021mft,Cipriani:2024ygw}. 

\subsection{Comparing with Leaver's continuous fraction method}

In this section, we briefly introduce the Leaver continuous fraction method \cite{Leaver:1985ax,Leaver:1990zz} and compare the QNMs obtained with the use of it to the analytic result for slowly damped modes \eqref{omegamatch}. 

The Frobenius exponents governing the leading behavior near the Fuchsian point $r = r_+$ are obtained from the ansatz
\be\label{frob}
R(r)\underset{r\to r_+}{\sim}(r-r_+)^\delta\,.
\ee
Plugging \eqref{frob} in \eqref{scawaeq} and expanding near $r = r_+$, we obtain
\be
\delta_\pm=\pm i\frac{r_+^2\omega}{r_+-r_-}\,,
\ee
where negative sign describes in-going behaviour at the horizon and viceversa.

The ansatz solving the equation \eqref{RNcan}, which guarantees outgoing waves at infinity and incoming waves at the horizon, is
\be
R_L(r)=e^{i\omega r}(r-r_+)^{-i\frac{r_+^2\omega}{r_+-r_-}}(r-r_-)^\rho\sum_{n=0}^\infty c_n\left(\frac{r-r_+}{r-r_-}\right)^n.
\ee
After inserting the previous ansatz inside the scalar wave equation \eqref{scawaeq}, $\rho$ can be chosen in order to minimize the terms of the recursion. The choice 
\be
\rho=-1+i\frac{(2r_+^2-r_-^2)\omega}{r_+-r_-}
\ee
leads to the recursion involving only three terms,
\be\label{3rec}
\alpha_n c_{n+1}+\beta_n c_n+\gamma_{n} c_{n-1}=0.
\ee
Here we defined
\begin{align}
\alpha_n&=(n+1)\left(1+n-\frac{2ir_+^2\omega}{r_+-r_-}\right),\nonumber\\
\beta_n&={-}\left(\ell{+}\frac{1}{2}\right)^2{-}2\left(n{+}\frac{1}{2}\right)^2{-}\frac{1}{4}{+}2i(1{+}2n)(2r_+{+}r_-)\omega+8(r_-^2{+}r_-r_+{+}r_+^2)\omega^2\nonumber\\
&+\frac{2r_-^2\omega(i+2i n+4r_-\omega)}{r_+-r_-},\nonumber\\
\gamma_n&=\frac{(n(r_--r_+)+2ir_+^2\omega)(n-2i(r_-+r_+)\omega)}{r_--r_+}.
\end{align}
The continuous fraction
\be\label{contfrac}
0=\beta_0-\frac{\alpha_0\gamma_1}{\beta_1-\frac{\alpha_1\gamma_2}{\beta_2-\frac{\alpha_2\gamma_3}{\beta_3-...}}}
\ee
can be solved numerically in $\omega$ in order to find the QNMs frequencies. In Tab. \ref{tabzero} we show some slowly damped modes, where we can see that the agreement between the Leaver method and the matching analytical estimation in \eqref{omegamatch} is achieved in the near extremal limit.
\begin{table}[]
\centering
\begin{tabular}{|c|c|c|}
\hline
$Q$      & Matching           & Leaver             \\ \hline
$0.5$    & $0. - {\rm i}0.497423 $  & $-$                \\ \hline
$0.9$    & $0. - {\rm i}0.422829 $  & $-$                \\ \hline
$0.99$   & $0. - {\rm i}0.216688 $  & $0.-{\rm i}0.299037 $       \\ \hline
$0.999$  & $0. - {\rm i}0.0819303 $ & $0. - {\rm i}0.0899565 $ \\ \hline
$0.9999$ & $0. - {\rm i}0.0275003 $ & $0. - {\rm i}0.0283005 $ \\ \hline
\end{tabular}
\caption{Zero damped modes for $M=1$, $\ell=1$ and $n=0$ for different values of Q. Since the matching procedure is expected to work efficiently only in the near-extremal limit, empty cells correspond to the regime where the matching procedure does not provide a reliable estimate of the frequency. In such cases, the Leaver method converges to a QNM frequency that is not zero-damped.}\label{tabzero}
\end{table}

\section{Scalar wave equation at first order in EFT corrections}
\label{sec:ZDMinEFT}
\subsection{Approximation of the wave equation by confluent Heun-type form}

The key point of our approach to finding slowly damped quasinormal modes is the use of an effective potential that includes the EFT corrections. The full expression (linearized in $g_4$ and $h_4$) is given by \eqref{QWnum}.

Although the equation with such a potential cannot be solved analytically, in the nearly extremal regime, we can find a reliable approximation by a Confluent Heun Equation (CHE), such that the solution would represent the known RN solution described in Section \ref{RNsec} with small corrections.

\begin{figure}
    \centering
    \includegraphics[width=0.49\linewidth]{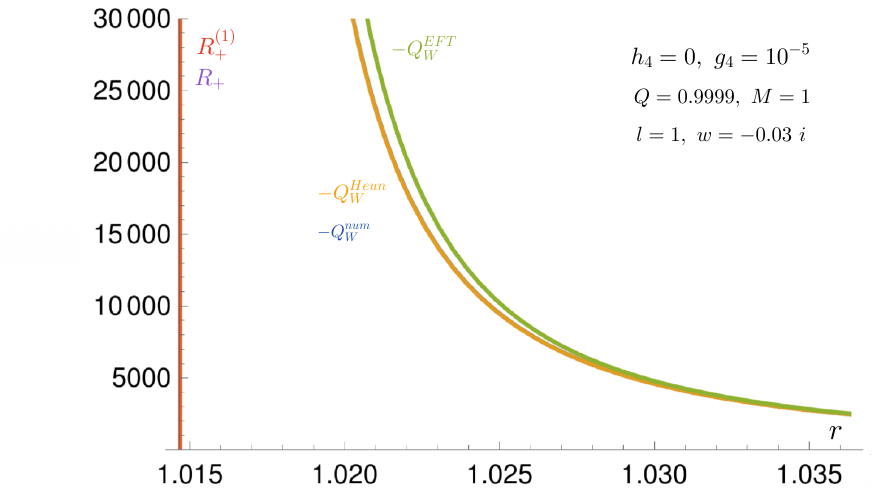}
    \includegraphics[width=0.49\linewidth]{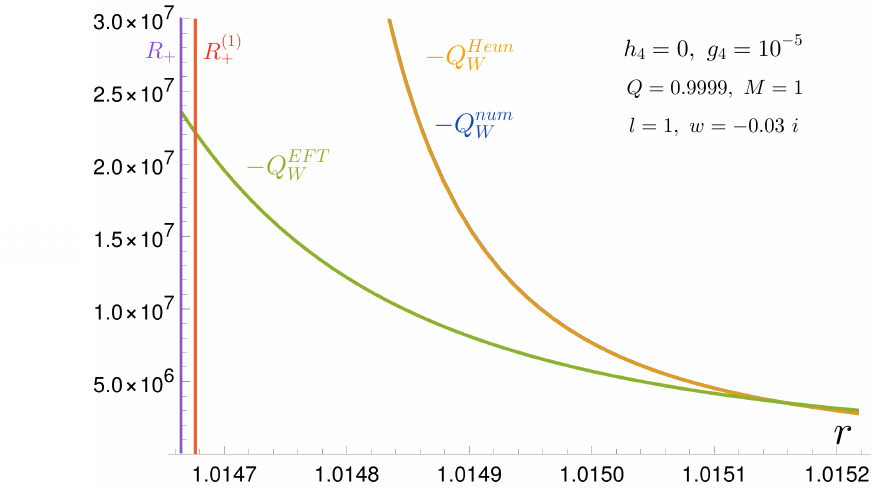}
   \includegraphics[width=0.45\linewidth]{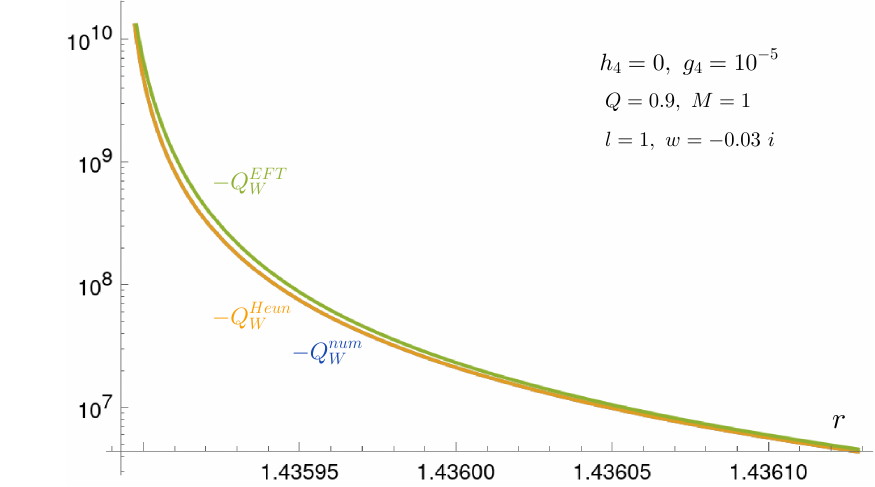}
    \includegraphics[width=0.45\linewidth]{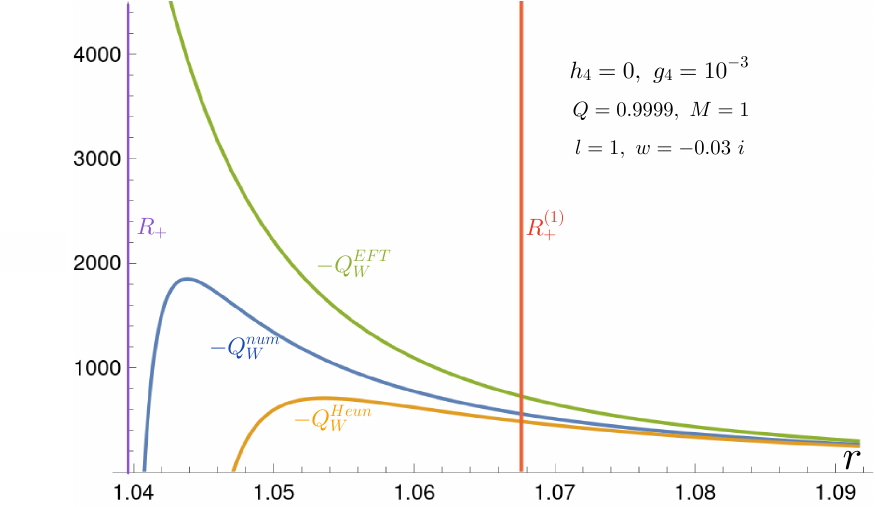}
    \caption{{\bf Exact and approximate effective potentials.} In all plots we show $Q_W^{num}$, $Q_W^{Heun}$, and $Q_W^{EFT}$ obtained from equations \eqref{QWnum}, \eqref{Qheun} and  \eqref{EFTQW}, respectively. The values of $R_+$ and $R_+^{(1)}$ correspond to the outer horizons obtained as an exact numerical root of $G(r)=0$, and the root $G(r)=0$ obtained approximately to the linear order in $g_4$ and given in \eqref{R1+-}. We choose $l=1,~\omega=-0.03 i$ because it corresponds to one of the slowly damped modes for the RN black hole, see Table \ref{tabzero}. The approximate location of the RN horizon corresponds to $r_+=1.01414$ for $Q=0.9999$, and $r_+=1.43589$ for $Q=0.9$. These values are less than $R_+$, so they are located to the left of the vertical axis. The upper two plots represent the case of small EFT coupling $g_4=10^{-5}$. From the upper left plot, it can be seen that the effective potential $Q_W^{EFT}$ is very close to the exact one for large $r$. The upper right plot shows zooming in on the values of $r$ close to the horizons, where $Q_W^{EFT}$ deviates from $Q_W^{num}$, while $Q_W^{Heun}$ still coincides with the exact potential. The lower left plot represents the case of $Q=0.9$, for which the approximation $Q_W^{EFT}$ works very well. The lower right plot shows the visible breakdown of the approximations for the larger value of $g_4=10^{-3}$ near the exact horizon $R_+$. The maxima of the effective potential are not shown because they correspond to very large values of $Q_W$ and appear to be extremely close to $R_+$ for the chosen parameters.}
    \label{plots}
\end{figure}

At this order in EFT corrections, the denominator of the effective potential has six roots, although only two of them ($R_+$ and $R_-$) are important in the nearly extremal regime. For this reason, we are separating them and keeping their values exact (i.e., we don't expand them in $g_4$ and $h_4$ at this stage). Thus, we can factorize the denominator in the following way,
\begin{equation}
\label{factorized}
\begin{split}
& -\frac{8 g_4 Q^4}{5 r^6} + h_4 \left(\frac{28 M Q^2}{r^5} - \frac{64 Q^4}{5 r^6} - \frac{16 Q^2}{r^4}\right) - \frac{2 M}{r} + \frac{Q^2}{r^2} + 1 =\\
& = \frac{\left(r-R_-\right) \left(r-R_+\right)}{r^6} \times \\
& \left(-\frac{8 g_4 \left(8 M^3 r^3 + Q^4 r (2 M - r) + 4 M Q^2 r^2 (M - r) + Q^6\right)}{5 Q^4} \right. \\
& \left. + \frac{4 h_4 \left(12 M^3 r^3 + Q^4 r (3 M - 4 r) + M Q^2 r^2 (6 M - 11 r) - 16 Q^6\right)}{5 Q^4} + r^4 \right).
\end{split}
\end{equation}
Recall that the values of roots $R_+$ and $R_-$ can also be approximated as
\begin{equation}
\label{Rm approx}
    R_-^{(1)}=M-\sqrt{M^2-Q^2}+\frac{4 g_4 Q^4-2 h_4 Q^2 \left(5 M \sqrt{M^2-Q^2}-5 M^2+4 Q^2\right)}{5 \left(\sqrt{M^2-Q^2}-M\right)^3 \left(M
   \sqrt{M^2-Q^2}-M^2+Q^2\right)},
\end{equation}
\begin{equation}
\label{Rp approx}
   R_+^{(1)}= M+\sqrt{M^2-Q^2}+\frac{4 g_4 Q^4+2 h_4 Q^2 \left(5 M \sqrt{M^2-Q^2}+5 M^2-4 Q^2\right)}{5 \left(\sqrt{M^2-Q^2}+M\right)^3 \left(M \sqrt{M^2-Q^2}+M^2-Q^2\right)}.
\end{equation}
However, we can keep them as exact expressions because they would provide a significantly better approximation for the effective potential. Our further strategy is to replace the denominator by \eqref{factorized} and expand the effective potential up to linear terms in $g_4$ and $h_4$. Schematically, we get an expression of the form
\begin{equation}
    Q_W^{{\rm num}}=\frac{A}{(r-R_+)^2}+\frac{B}{(r-R_+)}+\frac{C}{(r-R_-)^2}+\frac{D}{(r-R_-)}+E+\frac{F}{r}+\frac{G}{r^2}+O(r^{-3})+\dots.
\end{equation}
Coefficients $A,~B,~C,~D, \dots$ are quite long expressions containing the values of EFT couplings, exact positions of horizons $R_+,~R_-$ (they can be found from a numerical solution to the sextic equation), and parameters $l,~n,~ \omega$. It is important to note here that
\be 
A,C\propto(R_+-R_-)^{-2},~B,D\propto(R_+-R_-)^{-1},~F\propto (R_+-R_-),~G\propto(R_+-R_-)^2.
\ee
Such scaling in the extremal limit provides a hint that the terms $F,~G$ can be neglected. Indeed, we found that the terms containing negative powers of $r$ do not affect the form of the effective potential in the near extremal regime, see Figure \ref{plots}. The most relevant contributions are those which are singular near horizons $R+$ and $R_-$ (they are close to each other in the near extremal regime). Thus, if we keep only these terms, the equation becomes of Heun type in the near extremal limit, and, thus, admits an exact solution described in Section \ref{matchsec}. In order to obtain a confluent Heun equation, we keep an approximated value of the potential,
\begin{equation}\label{Qheun}
    Q_W^{{\rm Heun}}=\frac{A}{(r-R_+)^2}+\frac{B}{(r-R_+)}+\frac{C}{(r-R_-)^2}+\frac{D}{(r-R_-)}+E.
\end{equation}
Plots in Figure \ref{plots} demonstrate that this approximation is almost indistinguishable from the original potential \eqref{QWnum} in the near extremal regime with small enough EFT couplings.

If EFT couplings are small enough, we can also use the approximate values $R_+^{(1)}$, $R_-^{(1)}$, and obtain fully analytic expressions for slowly damped quasinormal modes with the first-order EFT corrections. In this approximation, the potential takes the form (only linear terms in EFT couplings are kept in this expression),
\begin{align}\label{EFTQW}
Q_W^{EFT}&=\frac{4 r^4 \omega ^2-4 \left(r-r_-\right)
   \left(r-r_+\right) \ell ^2-4 \left(r-r_-\right)
   \left(r-r_+\right) \ell
   +\left(r_--r_+\right){}^2}{4
   \left(r-r_-\right){}^2 \left(r-r_+\right){}^2}\nonumber\\
   &-g_4\frac{4  \left(4 r_+^4 \left(4 r_-^3+3 r_+
   r_-^2+2 r_+^2 r_-+r_+^3\right) \omega ^2-2
   r_-^5+r_+ r_-^4+r_+^5\right)}{5 r_-^2
   \left(r-r_+\right){}^2 \left(r_--r_+\right){}^2
   r_+^3}\nonumber\\
   &{+}h_4\frac{4r_+^4(23r_-^3{+}r_-^2r_+{-}r_-r_+^2{-}3r_+^3)\omega^2{+}6r_-^5{-}13r_-^4r_+{+}5r_-^3r_+^2{+}5r_-r_+^4{-}3r_+^5}{5 r_-^2(r{-}r_+)^2(r_-{-}r_+)^2r_+^3}.
\end{align}
Recall that here
\be 
r_+= M+\sqrt{M^2-Q^2},\quad r_-= M-\sqrt{M^2-Q^2}
\ee
are the positions of the RN horizons.

This potential can be mapped to the one studied in Section 3, such that the QNMs can be obtained analytically from the quantization condition for the Seiberg-Witten cycle \eqref{matching}. In Figure \ref{plots}, one can see that the approximation \eqref{EFTQW} is working well only for small EFT couplings, and cannot be applied in the extremal regime. However, there are also concerns about the very possibility of using the EFT description in the extremal regime, see the discussion in \cite{Horowitz:2022mly,Horowitz:2023xyl,Horowitz:2024dch,Horowitz:2024kcx}. Our expansion \eqref{EFTQW} holds only for the parameters within the EFT domain of validity, i.e.,
\be\label{validity}
\frac{g_{EFT}}{M^2-Q^2}\ll 1\footnote{Note that in this description the extremal case $M=Q$ cannot be considered, since the linear expansion of $r_+$ and $r_-$ in $g_{EFT}$ is not valid close to the extremal regime. This case would also correspond to a confluence of the radial differential equation.}\,,
\ee
 as in this limit the linear in $g_{EFT}$ approximation \eqref{R1+-} for the location of the external horizon is small.
 
%which is again mappable to \eqref{Q21def} in terms of the following dictionary
%\begin{align}
%u&=\left(\ell{+}\frac{1}{2}\right)^2{-}\left(r_-{+}r_+\right)^2 \omega^2{-}i
%   \left(r_-{-}r_+\right) \omega{+}\frac{4 g_4 \left(4 r_+^4 \left(4 r_-^3{+}3 r_+
%   r_-^2{+}2 r_+^2 r_-{+}r_+^3\right) \omega^2-2
%   r_-^5{+}r_+ r_-^4{+}r_+^5\right)}{5 r_-^2
%m_1&=\frac{i \left(r_-^2+r_+^2\right) \omega }{r_+-r_-}+\frac{g_4 \left(8 i r_+^4 \left(4 r_-^3+3 r_+
 %  r_-^2+2 r_+^2 r_-+r_+^3\right) \omega ^2-2 i
%   \left(2 r_-^5-r_+ r_-^4-r_+^5\right)\right)}{5
%   r_-^2 \left(r_--r_+\right) r_+^5 \omega }\nonumber\\
%m_2&=-i \left(r_-+r_+\right) \omega+\frac{g_4 \left(2 i \left(2 r_-^5-r_+
%   r_-^4-r_+^5\right)-8 i r_+^4 \left(4 r_-^3+3 r_+
%   r_-^2+2 r_+^2 r_-+r_+^3\right) \omega
%   ^2\right)}{5 r_-^2 \left(r_--r_+\right) r_+^5
%   \omega }\nonumber\\
%   m_3&=i\omega(r_++r_-)\quad,\quad q=2 i \left(r_+-r_-\right)
%\end{align}
\subsection{Matching the asymptotic expansion to Seiberg-Witten curve}\label{matchsec}
Applying the results outlined in section \ref{MAE}, the condition \eqref{condmatch} applied to the potential \eqref{EFTQW} provides the following analytic expression for the frequencies of the slowly damped modes 
\begin{equation}\label{omegaZDM}
\omega_{ZDM}=-\frac{i(\ell+n+1)}{Q^2}\sqrt{M^2-Q^2}-\frac{4i g_4(1+\ell+n)}{5Q^2\sqrt{M^2-Q^2}}-\frac{2ih_4(\ell+n+1) }{5Q^2\sqrt{M^2-Q^2}}+
\end{equation}
\begin{align}
&+\frac{i g_4 \sqrt{M^2{-}Q^2}}{5 Q^4}
   \Bigg[{-}\frac{8 \left((2 n{+}11)
   \ell {+}n (4 n{+}7){+}8 \ell
   ^2\right)}{2 \ell
   {+}1}{+}\frac{1}{n{+}\ell
   {+}1}{+}\frac{1}{(n{+}\ell
   {+}1)^3}{-}\frac{24}{2 \ell
   {+}1}\Bigg]\nonumber\\
   &+\frac{i h_4\sqrt{M^2-Q^2}}{10Q^4}\Bigg[\frac{4 (11-8 n) n}{2 \ell +1}+\ell  \left(\frac{184 n}{2 \ell
   +1}+68\right)+\frac{1}{n+\ell +1}+\frac{1}{(n+\ell +1)^3}\nonumber\\
   &+\frac{4}{2 \ell
   +1}+72\Bigg].
\end{align}

This expression can be trusted only when \eqref{validity} is satisfied. It is interesting to mention that in the leading order of the extremality parameter $\sqrt{M^2-Q^2}$ we obtain the correction proportional to the combination $h_4+2 g_4$. The same combination emerges as an implication of the WGC, requiring it to be positive, see Eq. \eqref{WGCcond}. In addition, the QNM causality requirement recently proposed in \cite{Melville:2024zjq},
\be 
\label{causality}
\delta(-{\rm Im\,}\omega)>0,
\ee
leads to the same statement for near extremal RN black holes, as it follows from our result \eqref{omegaZDM},
\be 
\delta(-{\rm Im\,}\omega)\approx h_4+2 g_4>0.
\ee
Here $\delta(-{\rm Im\,}\omega)$ is an EFT correction to the QNM frequencies. The statement requires that the EFT couplings compatible with causality should make the damping rate of QNMs larger compared to the GR case. Thus, our computation shows the complete coincidence between WGC and QNM causality requirements for the scalar wave in near extremal RN black hole geometry.

A QNM causality, as it is formulated in \cite{Melville:2024zjq}, requires the difference between the characteristic lifetimes of the QNMs to be resolvable in the sense of the time-energy uncertainty principle in quantum mechanics. This imposes a condition allowing for a small violation of \eqref{causality} containing the real part of the QNM frequency, effectively playing the role of energy. However, the ZDM frequencies have zero real parts, and the resolvability condition should be formulated in a different way, as they are not waves. In particular, it is not fully clear which parameter has a physical meaning of the energy of the corresponding scalar field configuration, and how the quantum mechanical uncertainty in the measurement of the lifetime of the perturbation should be properly estimated. We leave with a better understanding of these fundamental questions for future work.

Our result aligns well with the growing evidence that different definitions of causality and EFT consistency are related to each other. In particular, causality probes including time delays of the wave propagation on top of the background \cite{Camanho:2014apa, deRham:2021bll,Chen:2021bvg, Chen:2023rar,CarrilloGonzalez:2023emp,Nie:2024pby,CarrilloGonzalez:2025fqq} are in many cases showing very similar constraints as positivity bounds from the scattering amplitudes \cite{CarrilloGonzalez:2022fwg,deRham:2023ngf}. Although the relation between the analyticity of the scattering amplitudes and the absence of time advances is not direct and obvious \cite{Toll:1956cya}, the resulting bounds have a similar form, even though the setups look different.

%Since in this scheme $\ell>n$ is intended, the imaginary part of $\omega_{ZDM}$ is negative in the range
%\be
%0<g_4<\frac{5 r_+ \left(r_+-r_-\right) (2 \ell +1)
%   (n+\ell +1)^2}{24 (\ell -n)}\,.
%\ee
%So it seems that if $r_-$ is sufficiently close to $r_+$, some instability could arise.

%As an example, let me choose $M=1$, $Q=0.99999$, $l=1$, $n=0$,$g_4=0.03$. QNMs frequencies are:
%\be
%\omega_{ZDM}={\rm i} 0.00297565\quad,\quad \omega_{Leaver}={\rm i}0.0330605
%\ee
%$\omega_{Leaver}$ was first computed using 400 iterations in the continued fraction, and then again with 500 nested terms. The result of the numerical root search did not change with increased precision, indicating convergence.

\section{EFT corrections to prompt ringdown modes}
\label{sec:PRM}
In this Section, we obtain the results for the prompt ringdown  modes using such methods as WKB approximation \cite{Iyer:1986vv}, Leaver \cite{Leaver:1985ax,Leaver:1990zz}, and numerical integration of the wave equation. 

Even though one could consider higher-order terms in the WKB expansion \cite{PhysRevD.35.3621,Konoplya:2003ii,Konoplya:2019hlu}, we limit our discussion to the leading-order WKB approximation. This (not too) rough estimate will serve as a seed for the numerical search of roots appearing in Leaver's continuous fraction method and in numerical integration. Since the geometry under examination is a spherically symmetric, asymptotically flat black hole, the eikonal limit based on the geodesic approach is very effective, although it is known that in some cases it may fail \cite{Konoplya:2017wot}.

We provide the tables of the QNM frequencies and show that the results obtained by different methods coincide.

\subsection{Geodesics and WKB approximation}

We start with studying the motion of a scalar massless particle in the geometry \eqref{EFTsol1},\eqref{EFTsol2} in Hamiltonian formalism,
\be\label{hamilton}
\mathcal{H}=g^{\m\n}P_\m P_\n=0\quad,\quad P_\m=\frac{\partial \mathcal{L}}{2\partial \dot{x}^\m}.
\ee
Here $\mathcal{L}$ is the Lagrangian, and dot represents the derivative with respect to the proper time.

Similarly to the well-known case of RN geometry, $P_t$ and $P_\phi$ are the constants of the motion and can be interpreted as the energy and the angular momentum of a particle,
\be
P_t=-E=-G(r)\dot{t},\quad P_r=\frac{\dot{r}}{G(r)},\quad P_\q^2=r^2\dot{\q},\quad P_{\phi}=J=r^2\sin^2\q \dot{\phi}.
\ee
Thus, the Hamilton mass shell condition \eqref{hamilton} in terms of the conserved quantities can be rewritten as
\be
-\frac{E^2}{G(r)}+G(r)P_r^2+\frac{P_\q}{r^2}+\frac{J^2}{r^2\sin^2\q}=0.
\ee
The radial and angular dynamics can be easily separated by the introduction of the Carter constant \cite{Chandrasekhar:1984siy}
\begin{align}
P_r^2&=Q_R(r)=\frac{E^2}{G^2(r)}-\frac{K^2}{r^2 G(r)} \nn\\
     P_\q^2&=Q_A(\q)=K^2-\frac{J^2}{\sin^2\q}.
\end{align}

Due to the spherical symmetry, we can study the equatorial motion without loss of generality, since the motion is always planar. Setting $\theta=\pi/2$, we obtain $P_\q=0$, so that $K=J$. If we introduce the impact parameter $b=J/E$, the photon-spheres are defined as the double zeros of the radial potential for geodesics $Q_R$, or, equivalently, as the location where both radial velocity and acceleration are vanishing,
\be\label{photonsphere}
Q_R(r_c,b_c)=Q_R'(r_c,b_c)=0.
\ee
Here $'$ means derivatives w.r.t. $r$. Unfortunately, due to the sixth-degree algebraic equation coming from the function $G(r)$ in \eqref{EFTsol1}, the condition for the photon-spheres \eqref{photonsphere} can be solved only numerically.

\begin{figure}
    \centering
    \includegraphics[width=0.5\linewidth]{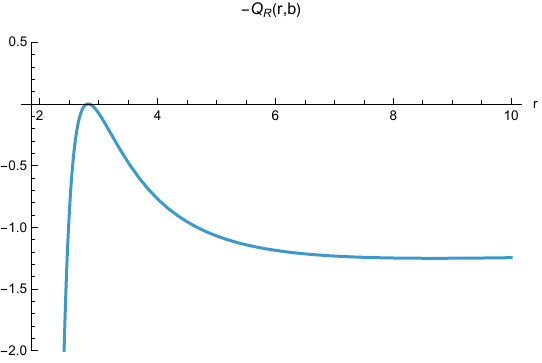}
    \caption{Effective potential in the critical regime given as parameters $M=1$, $Q=0.5$, $g_4=0.01$. The horizon is at $r_+=1.86607$, the critical unstable radius of the circular photon sphere is at $r_c=2.8229$, corresponding to a critical impact parameter $b_c=4.96793$.}
 \label{potencrit}
\end{figure}

In the eikonal approximation, the real and imaginary parts of the QNM frequencies are consistent with the prompt ringdown modes. These modes are associated with the unstable light ring. Because of the shape of the potential, a wave impinging on the compact object is partially reflected by the unstable photon sphere, so these modes constitute the first signal detected by an observer at infinity. In eikonal approximation, prompt ring-down modes can be expressed as \cite{Bianchi:2020des,Cardoso:2008bp}
\be
\omega_{QNM}\sim E_c-{\rm i}(2n+1)\lambda\quad ,\quad E_c=\frac{\ell}{b_c},
\ee
where $b_c$ is the critical impact parameter of the unstable circular orbit forming the light ring (see Fig.\ref{potencrit}) while $\lambda$ is the Lyapunov exponent governing the chaotic behavior of the geodesic motion near the photon-sphere. The nearly critical geodesics fall with radial velocity, 
\be
\frac{dr}{dt}\sim -2\lambda (r-r_c).
\ee
where
\be\label{Lyapunove}
\lambda=\left(\sqrt{2}\partial_E Q_R(r_c,E_c)\right)^{-1}\sqrt{\partial_r^2 Q_R(r_c,E_c)}.
\ee

In the classically allowed regions where $Q_W(r,\omega)$ in \eqref{scalwaveeq} is positive and large, the wave equation can be solved in a semiclassical WKB approximation,
\be\label{waveWKB}
\psi(r)=\frac{1}{\sqrt[4]{Q_W(r,\omega)}}{\rm exp}\left(\pm {\rm i \int^r\sqrt{Q_W(r',\omega)}}dr'\right)\,.
\ee
This approximation fails near the zeros of $Q_W(r,\omega)$ called $r_\pm$, which are the turning points of the classical motion. Thus, the matching with the allowed solutions in the classically forbidden region 
\be\label{decWKB}
\psi(r)=\frac{1}{\sqrt[4]{-Q_W(r,\omega)}}{\rm exp}\left(\pm {\rm  \int^r\sqrt{-Q_W(r',\omega)}}dr'\right)
\ee
is achieved by linearizing the effective potential in the vicinity of the turning points and connecting the solutions \eqref{waveWKB} and \eqref{decWKB} by using the Airy functions \cite{Bena:2019azk}. These matching procedures imply the Bohr-Sommerfeld (BS) quantization condition,
\be\label{BSq1}
\int_{r_-}^{r_+}\sqrt{Q_W(r,\omega)}dr=\pi\left(n+\frac{1}{2}\right),
\ee
where $n$ is a non-negative integer also known as the overtone number. When the two turning points are almost coincident (so when we are near the critical geodesic), the BS condition can be approximated as follows,
\be\label{BSq2}
\int_{r_-}^{r_+}\sqrt{Q_{W}(r,\omega)}dr{\sim}\int_{r_-}^{r_+}\sqrt{Q_W(r_c,\omega){+}\frac{Q_W''(r_c,\omega)}{2}(r{-}r_c)^2}dr{\sim}\frac{i\pi Q_W(r_c,\omega)}{\sqrt{2Q_W''(r_c,\omega)}}
\ee
since $Q_W'(r_c,\omega)=0$. The frequencies acquire an imaginary part $\omega=\omega_R+{\rm i}\omega_I$ and represent the QNMs frequencies.
It turns out that in this WKB approximation, the real part of the QNMs frequencies assumes exactly the critical value $\omega_c$, which should be considered much larger than the imaginary part. 

The results for the QNM frequencies obtained from the BS quantization condition \eqref{BSq1}, \eqref{BSq2} are collected in Tab. \ref{tabqnm3}, \ref{tabqnm4}.

The eikonal limit is obtained after replacing
\begin{equation}
\ell=\frac{J}{\hbar}-\frac{1}{2}\quad,\quad \omega=\frac{E}{\hbar}\quad,\quad \psi(r)\sim e^{i\frac{S_0(r)}{\hbar}}\,.
\end{equation}
so that the potential appears to be
\begin{align}
   \hbar^2 \left(\frac{S_0(r)}{dr}\right)^2=Q_{geo}(r,E)&=\frac{E^2 r^4-J^2(r-r_-)(r-r_+)}{(r-r_+)^2(r-r_-)^2}\nonumber\\
    &-g_4\frac{16r_+(4r_-^3+3r_-^2r_++2r_- r_+^2+r_+^3)E^2}{5r_-^2(r-r_+)^2(r_+-r_-)^2}\nonumber\\
    &-h_4\frac{4r_+(23r_-^3+r_-^2r_+-r_-r_+^2-3r_+^3)E^2}{5r_-^2(r-r_+)^2(r_+-r_-)^2}.
\end{align}
The critical conditions
\begin{equation}
Q_{geo}(r_c,J_c)=\frac{dQ_{geo}(r_c,J_c)}{dr}=0
\end{equation}
can be solved perturbatively in the EFT coupling as follows,
\begin{align}
r_c&=\frac{1}{4} \left(\sigma {+}3
   \left(r_-{+}r_+\right)\right)\nonumber\\
   &+g_4\frac{32 \left(4 r_-^3{+}3 r_+ r_-^2{+}2 r_+^2
   r_-{+}r_+^3\right) \left({-}r_- \left(\sigma {+}7
   r_+\right){+}2 r_+ \left(\sigma {+}3 r_+\right){+}3
   r_-^2\right)}{5 \sigma  r_-^2
   \left(r_-{-}r_+\right) \left(\sigma {+}3
   \left(r_+{+}r_-\right)\right)^2}\nonumber\\
   &+h_4\frac{8(23r_-^3+r_-^2r_+-r_-r_+^2-3r_+^3)(3r_-^2+2r_+(3r_++\sigma)-r_-(7r_++\sigma))}{5r_-^2\sigma(r_--r_+)(\sigma+3(r_++r_-))^2}\nonumber\\
J_c&=\frac{E \left(\sigma {+}3
   \left(r_-{+}r_+\right)\right)^{3/2}}{2 \sqrt{2}
   \sqrt{\sigma {+}r_-{+}r_+}}{-}g_4\frac{16 \sqrt{2} E \left(4 r_-^3{+}3 r_+ r_-^2{+}2
   r_+^2 r_-{+}r_+^3\right)}{{5 \sigma
    r_-^2 \left(r_-{-}r_+\right)^2 \left(\sigma
   {+}r_-{+}r_+\right)^{3/2} \left(\sigma {+}3
   \left(r_-{+}r_+\right)\right)^{3/2}}}\nonumber\\
   &\times\Big[ \sigma ^2 \left(2
   r_-^2{-}7 r_+ r_-{+}7 r_+^2\right){-}2 \sigma 
   \left(r_-{-}3 r_+\right) \left(3 r_-^2{-}5 r_+ r_-{+}4
   r_+^2\right)\nonumber\\
   &+r_+ \left(r_-{+}r_+\right) \left(9
   r_-^2{-}14 r_+ r_-+9 r_+^2\right)\Big]\nonumber\\
   &{-}h_4\frac{4 \sqrt{2} E \left(23 r_-^3{+}r_+ r_-^2{-}r_+^2 r_-{-}3 r_+^3\right)}{5 \sigma 
   r_-^2 \left(r_-{-}r_+\right)^2 \left(\sigma {+}r_-{+}r_+\right)^{3/2}
   \left(\sigma {+}3 \left(r_-{+}r_+\right)\right)^{3/2}}\times\Bigg[\sigma ^2 \left(2 r_-^2{-}7 r_+ r_-{+}7 r_+^2\right)\nonumber\\
   &-2 \sigma  \left(r_--3 r_+\right)
   \left(3 r_-^2-5 r_+ r_-+4 r_+^2\right)+r_+ \left(r_-+r_+\right) \left(9
   r_-^2-14 r_+ r_-+9 r_+^2\right)\Bigg]
\end{align}
where we defined 
\be
\sigma=\sqrt{9r_+^2+9r_-^2-14r_+r_-}\,.
\ee
Using \eqref{Lyapunove}, we can compute the Lyapunov exponent
\be
\lambda=\lambda_{RN}+g_4 \lambda_{g_4}+h_4 \lambda_{h_4}
\ee
\begin{eqnarray}
\label{Lyapunov}
\lambda_{RN} &=&   \frac{4 \sqrt{2} \rho  \left(3 r_--r_++\sigma \right)^2 \left(-r_-+3 r_++\sigma \right)^2}{\sqrt{r_-+r_++\sigma } \left(3 \left(r_-+r_+\right)+\sigma \right)^4 \left(r_-
   \left(\sigma -2 r_+\right)+r_+ \left(3 r_++\sigma \right)+3 r_-^2\right)^2}\nonumber\\
\lambda_{g_4}&=&\frac{1}{5 \rho  r_-^2 \left(r_-{-}r_+\right)^2 \sigma  \left(3
   r_-{-}r_+{+}\sigma \right)^2 \left(r_-{+}r_+{+}\sigma \right)^{3/2} \left({-}r_-{+}3
   r_+{+}\sigma \right)^2 \left(3 \left(r_-{+}r_+\right){+}\sigma \right)^6}\nonumber\\
   &\times&\Big[131072 \sqrt{2} r_+ \left(4 r_-^3+3 r_+ r_-^2+2 r_+^2 r_-+r_+^3\right)
   \left(r_- \left(\sigma -2 r_+\right)+r_+ \left(3 r_++\sigma \right)+3
   r_-^2\right)^2\nonumber\\
   &\times&\left(r_-^8 \left(9 \sigma -21 r_+\right)+12 r_+^2 r_-^6
   \left(9 r_++\sigma \right)-6 r_+^3 r_-^5 \left(69 r_++4 \sigma \right)+42 r_+^4
   r_-^4 \left(25 r_++3 \sigma \right)\right.\nonumber\\
   &{-}&\left.12 r_+^5 r_-^3 \left(129 r_+{+}20 \sigma
   \right){+}4 r_+^6 r_-^2 \left(375 r_+{+}79 \sigma \right){-}27 r_+^7 r_- \left(31
   r_+{+}8 \sigma \right){+}81 r_+^8 \left(3 r_+{+}\sigma \right)\right.\nonumber\\
   &+&\left.27 r_-^9+20 r_+^2
   r_-^7\right)\Big]\nonumber\\
\lambda_{h_4}&=&\lambda_{g_4}\times\frac{23r_-^3+r_-^2r_+-r_-r_+^2-3r_+^3}{4(4r_-^3+3r_-^2r_++2r_-r_+^2+r_+^3)}
\end{eqnarray}
where
\begin{eqnarray}
    \rho^2&=&81 r_-^6 \left(\sigma {-}3 r_+\right){+}3 r_+ r_-^5 \left(77 r_+{-}6 \sigma
   \right){+}r_+^2 r_-^4 \left(25 r_+{+}47 \sigma \right){+}r_+^3 r_-^3 \left(25 r_+{+}36
   \sigma \right)\nonumber\\
   &+&r_+^4 r_-^2 \left(231 r_++47 \sigma \right)-9 r_+^5 r_- \left(27
   r_++2 \sigma \right)+81 r_+^6 \left(3 r_++\sigma \right)+243 r_-^7
\end{eqnarray}
The QNM causality condition \eqref{causality} requires \cite{Melville:2024zjq} 
\be
g_4 \lambda_{g_4}+h_4 \lambda_{h_4}>0,
\ee
which translates to 
\be
2 g_4+f\left(\frac{r_-}{r_+}\right)h_4>0,
\ee
where for $0<\gamma<1$ we have
\be
f(\gamma)=\frac{23 \gamma ^3+\gamma ^2-\gamma -3}{2 \left(4 \gamma ^3+3 \gamma ^2+2 \gamma +1\right)}, \quad -\frac{3}{2}<f(\gamma)<1.
\ee.
However, both limits $r_-\rightarrow 0$ and $r_-\rightarrow r_+$ represent the situations where the expansion \eqref{Lyapunov} are not applicable. For this reason, strictly speaking, we cannot make a robust conclusion that
\be 
-2 g_4<h_4<\frac{4}{3} g_4.
\ee
The most optimal constraints obtained from this method on both sides correspond to the choice of $r_-$ for which the method cannot be applied. Thus, the corrections to Lyapunov exponent provide a weaker statement than the one derived in Section 4 from zero damping modes.

\subsection{Leaver method}

The leading behaviors on the horizon are:
\be\label{rpbc}
R(r)\underset{r\to r_+}{\sim}(r-r_+)^{\alpha_{\pm}},
\ee
where the Frobenius coefficient is
\begin{align*}
    \alpha_{\pm}&=\pm\frac{\sqrt{P_0+g_4 P_{g_4}+h_4 P_{h_4}}}{\sqrt{5}r_+^{3/2}r_-(r_+-r_-)},\nonumber\\
    P_0&=-5r_+^7r_-^2\omega^2,\nonumber\\
    P_{h_4}&=4 r_+^4 \left(23 r_-^3{+}r_+ r_-^2{-}r_+^2 r_-{-}3 r_+^3\right) \omega
   ^2{+}6 r_-^5{-}13 r_+ r_-^4{+}5 r_+^2 r_-^3{+}5 r_+^4 r_-{-}3 r_+^5,\nonumber\\
   P_{g_4}&=4  \left(4 r_+^4 \left(4 r_-^3+3 r_+ r_-^2+2 r_+^2 r_-+r_+^3\right)
   \omega ^2-2 r_-^5+r_+ r_-^4+r_+^5\right).
\end{align*}

   Here $\alpha_-$ encodes the correct ingoing boundary condition at the horizon, while the leading behaviors at infinity can be written as
   \be
R(r)\underset{r\to\infty}{\sim}e^{\pm i\omega r}\,.
   \ee
   Thus, we can use the following ansatz for the Leaver continuous fraction procedure,
   \be\label{Lans}
R(r)=e^{i\omega r}(r-r_+)^{\alpha_-}(r-r_-)^\beta\sum_{n=0}^\infty c_n\left(\frac{r-r_+}{r-r_-}\right)^n.
\ee
Here, the exponent $\beta$ is chosen in order to ensure the arising of a three-term recursion relation
\be
\beta=-1+i\omega(r_++r_-)-\alpha_-
\ee
Performing a change of variable 
\be
z=\frac{r-r_+}{r-r_-}
\ee
and plugging \eqref{Lans} in \eqref{EFTQW} we obtain a three terms recursion of the form \eqref{3rec} with the extra condition $c_{-1}=0$. This recursion relation can be solved by the continuous fraction \eqref{contfrac}.
The coefficients of the recursion are:
\begin{align}\label{LeaverRNrootscoeff}
\alpha_n&=(n+1)(n+1+2\alpha_-)\,,\nonumber\\
\beta_n&={-}\ell(\ell{+}1){-}1{-}2n(n{+}1){-}2\alpha_-^2{+}2ir_+\omega(1{+}2n){+}\frac{2r_+^3(r_+{-}2r_-)\omega^2}{(r_+{-}r_-)^2}{+}2\alpha_-(2ir_+\omega{-}2n{-}1)\,,\nonumber\\
\gamma_n&=n^2-2i n(r_++r_-)\omega+\alpha_-^2+\frac{r_+^2(2r_-^2-r_+^2)\omega^2}{(r_+-r_-)^2}+2\alpha_-(n-i(r_++r_-)\omega).
\end{align}

Now we focus our attention on the exponent $\alpha_\pm$ in \eqref{rpbc}. Let us replace $\omega=\omega_R+i \omega_I$ knowing that in the case of stable modes $\omega_I=-|\omega_I|$. In order to ensure purely ingoing boundary conditions at the horizon, we impose that the imaginary part of the radicand must be positive. Such a condition provides an expression
\be\label{constr2}
5r_-^2r_+^3{-}4h_4(23r_-^3{+}r_-^2r_+{-}r_- r_+^2{-}3r_+^3){-}16g_4(4r_-^3{+}3r_-^2r_+{+}2r_-r_+^2{+}r_+^3)>0\,.
\ee

Some ZDMs computed using this approximation are displayed in the third column of Table \ref{LeaverRNroots}. The comparison is done with \eqref{omegaZDM} (first column) and with the Leaver method implemented in the solution with the exact roots \eqref{Qheun} (second column).

\subsection{Numerical Integration method}
In this subsection, we briefly describe the numerical procedure implemented in Mathematica used for the QNMs computation. The starting point consists of finding the leading and a sufficient number of subleading terms at infinity and at the horizon. At infinity, the radial wave function behaves as
\be\label{solinf}
R_\infty(r)\simeq e^{{\rm i\omega r}}r^{2{\rm i}M \omega}\sum_{n=0}^{N_\infty}c_n r^{-n} \,.
\ee
The behavior at infinity can be captured by
\be\label{solh}
R_H(r)\simeq (r-r_H)^\alpha \sum_{n=0}^{N_H} d_n(r-r_H),
\ee
where $\alpha$ approaches $-2{\rm i}M \omega$ in the Schwarzschild limit. 

The numerical integration is performed starting from the horizon and proceeding up to infinity using as boundary conditions \eqref{solinf} and \eqref{solh} and their first derivatives. Fixing the unconstrained coefficients $c_0=d_0=1$, we can construct the numerical Wronskian whose zeros can be interpreted as the QNMs frequencies. In Tables \ref{tabqnm3}, \ref{tabqnm4}, we show some results valid for the first overtone number $n=0$, which are obtained by fixing $N_H=N_\infty=10$. It is known that this type of numerical algorithm is not very efficient for highly-damped modes
\cite{Cardoso:2014sna,Cipriani:2024ygw,Bena:2024hoh}. For example, with the increased precision $N_H=15$ and $N_\infty=30$, the modes with overtone number $n=1$ still cannot be computed without numerical issues. However, in principle, with a sufficient number of subleading terms, modes with overtone greater than zero can also be found.

\subsection{Tables of QNM frequencies}

In all tables of this section, we set $h_4=0$, as it doesn't affect the applicability of the discussed methods if it is assumed to be the same order of magnitude as $g_4$.

\begin{table}[]
\centering
\begin{tabular}{|c|c|c|c|}
\hline
$Q$ & $\omega_{ZDM}$ & Leaver (exact roots) & Leaver (RN roots)\\ \hline
$0.9$  & $-{\rm i}1.07639$            & $-{\rm i}1.47217$ & $-{\rm i}1.47195$    \\ \hline
$0.99$  & $-{\rm i}0.287996$            & $-{\rm i}0.299177$ & $-{\rm i}0.299041$     \\ \hline
$0.999$  & $-{\rm i}0.0899633$            & $-{\rm i}0.0903221$ & $-{\rm i}0.0899557$    \\ \hline
$0.9999$  & $-{\rm i}0.0294225$            & $-{\rm i}0.0294129$ & $-{\rm i}0.0282977$     \\ \hline
\end{tabular}
\caption{$\ell=1$, $g_4=10^{-5}$, $M=1$, lowest overtone. We compare results provided by \eqref{omegaZDM} (first column) with the Leaver method implemented on the CHE-like solution with exact roots \eqref{Qheun} (second column and RN roots \eqref{LeaverRNrootscoeff} (third column). }\label{LeaverRNroots}
\end{table}

\begin{table}[]
\centering
\begin{tabular}{|c|c|c|c|}
\hline
$Q$ & WKB& Numerical (non-expanded) & Numerical (CHE-like expanded) \\ \hline
$0.6$ &$0.274006 - {\rm i}0.106959$ & $0.313527 - {\rm i}0.0991528$    & $0.265033 - {\rm i}0.0850159$         \\ \hline
$0.7$ & $0.283844 - {\rm i}0.10693$ & $0.32277 - {\rm i}0.099352$    & $0.29511 - {\rm i}0.090372$         \\ \hline
$0.8$ & $0.297506 - {\rm i}0.106128$ & $0.335211 - {\rm i}0.09905$    & $0.35988 - {\rm i}0.08782$          \\ \hline
$0.9$ & $0.31775 - {\rm i}0.102985$ & $0.352581 - {\rm i}0.0972046$    & $0.355235 - {\rm i}0.0965532$          \\ \hline
$0.95$ & $0.332434 - {\rm i}0.0988134$ & $0.364045 - {\rm i}0.0946749$    & $0.36415 - {\rm i}0.0951443$          \\ \hline
$0.99$ & $0.348478 - {\rm i}0.091618$ & $0.374818 - {\rm i}0.0907836$    & $0.374094 - {\rm i}0.0914614$          \\ \hline
$0.999$ & $0.352962 - {\rm i}0.0889113$ & $0.377392 - {\rm i}0.0895647$ & $0.37659 - {\rm i}0.0901418$\\\hline
\end{tabular}
\caption{\label{tab4}$\ell=1$, $M=1$, $g_4=0.001$, lowest overtone. In the first two columns, WKB and numerical integration have been implemented on the non-expanded canonical wave equation \eqref{QWnum}. In the third column, the numerical integration has been implemented on the approximate CHE-like solution with exact roots \eqref{Qheun}.}
\end{table}

\begin{table}[]
\centering
\begin{tabular}{|c|c|c|c|}
\hline
$\ell$ & Eikonal & WKB & Numerical                \\ \hline
$0$    & $-$      & $-$  & $0.134055 - {\rm i}0.0950224 $ \\ \hline
$1$    & $0.249876 - {\rm i}0.088953 $      & $0.352905 -{\rm i}0.0901165$   & $0.378031 - {\rm i}0.0897661 $ \\ \hline
$2$    & $0.499752 - {\rm i}0.088953 $      & $0.611793 - {\rm i}0.0893476 $  & $0.626624 - {\rm i}0.0892448 $ \\ \hline
$3$    & $0.749628 - {\rm i}0.088953 $      & $0.8654 - {\rm i}0.0891512 $  & $0.875946 - {\rm i}0.0891012 $ \\ \hline
$4$    & $0.999504 - {\rm i}0.088953 $      & $1.11733 - {\rm i}0.0890721 $  & $1.12552 - {\rm i}0.0890424 $  \\ \hline
$5$    & $1.24938 - {\rm i}0.088953 $      & $1.3685 - {\rm i}0.0890325 $  & $1.3752 - {\rm i}0.0890127 $   \\ \hline
$6$    & $1.49926 - {\rm i}0.088953 $      & $1.61928 - {\rm i}0.0890098 $  & $1.62494 - {\rm i}0.0889957 $  \\ \hline
$7$    & $1.74913 - {\rm i}0.088953 $      & $1.86981 - {\rm i}0.0889956$  & $1.87471 - {\rm i}0.0889851 $  \\ \hline
$8$    & $1.99901 - {\rm i}0.088953 $      & $2.12019 - {\rm i}0.0889862 $  & $2.12451 - {\rm i}0.088978 $   \\ \hline
$9$    & $2.24888 - {\rm i}0.088953 $      & $2.37046 - {\rm i}0.0889795 $  & $2.37433 - {\rm i}0.088973 $   \\ \hline
$10$   & $2.49876 - {\rm i}0.088953 $      & $2.62066 - {\rm i}0.0889747 $  & $2.62416 - {\rm i}0.0889693 $  \\ \hline
\end{tabular}
\caption{\label{tabqnm3}Scalar QNMs frequencies with parameters $M=1$, $Q=1$, $g_4=0.01$ corresponding to various angular quantum number $\ell$ and ad fixed overtone number $n=0$. WKB and numerical integration have been implemented on the non-expanded wave equation \eqref{QWnum}. Eikonal and WKB become reliable approximations for QNMs frequencies if $\ell>>n$. For this reason, empty cells are left in the first entries of the first two columns.}
\end{table}

\begin{table}[]
\centering
\begin{tabular}{|c|c|c|c|}
\hline
$\ell$ & Eikonal & WKB & Numerical                \\ \hline
$0$    & $-$      & $-$  & $0.142158 - {\rm i}0.100662 $ \\ \hline
$1$    & $0.248835 - {\rm i}0.0933079 $       &  $0.348027 - {\rm i}0.101178 $  & $0.379992 - {\rm i}0.0947284 $ \\ \hline
$2$    & $0.497671 - {\rm i}0.0933079 $      & $0.60712 - {\rm i}0.096244 $   & $0.626216 - {\rm i}0.0938125 $ \\ \hline
$3$    & $0.746506 - {\rm i}0.0933079 $       & $0.860262 - {\rm i}0.0948227 $   & $0.873875 - {\rm i}0.0935541 $ \\ \hline
$4$    & $0.995342 - {\rm i}0.0933079 $       & $1.11148 - {\rm i}0.0942286 $   & $1.12205 - {\rm i}0.0934523 $  \\ \hline
$5$    & $1.24418 - {\rm i}0.0933079 $       & $1.36182 - {\rm i}0.0939257 $   & $1.37047 - {\rm i}0.0934027 $   \\ \hline
$6$    & $1.49301 - {\rm i}0.0933079 $       & $1.6117 - {\rm i}0.0937509 $   & $1.61901 - {\rm i}0.0933749 $  \\ \hline
$7$    & $1.74185 - {\rm i}0.0933079 $       & $1.8613 - {\rm i}0.0936409 $   & $1.86764 - {\rm i}0.0933578 $  \\ \hline
$8$    & $1.99068 - {\rm i}0.0933079 $       & $2.11072 - {\rm i}0.0935673 $   & $2.11631 - {\rm i}0.0933465 $   \\ \hline
$9$    & $2.23952 - {\rm i}0.0933079 $       & $2.36002 - {\rm i}0.0935157 $   & $2.36502 - {\rm i}0.0933387 $   \\ \hline
$10$   & $2.48835 - {\rm i}0.0933079 $       & $2.60923 - {\rm i}0.093478 $   & $2.61375 - {\rm i}0.093333 $  \\ \hline
\end{tabular}
\caption{\label{tabqnm4}Scalar QNMs frequencies with parameters $M=1$, $Q=1$, $g_4=0.1$ corresponding to various angular quantum number $\ell$ and ad fixed overtone number $n=0$. WKB and numerical integration have been implemented on the non-expanded wave equation \eqref{QWnum}. Eikonal and WKB become reliable approximations for QNMs frequencies if $\ell>>n$. For this reason, empty cells are left in the first entries of the first two columns.}
\end{table}

%\section{Gravitational and electromagnetic perturbations}
%Let's compute at the linearized level the gravitational and electromagnetic perturbations by replacing
%\be
%g_{\m\n}\to g_{\m\n}+h_{\m\n}\quad,\quad F_{\m\n}\to F_{\m\n}+f_{\m\n}
%\ee
%where as a consequence
%\be
%g^{\m\n}\to g^{\m\n}-h^{\m\n}\quad,\quad \sqrt{-g}\to \sqrt{-g}\left(1+\frac{1}{2}g^{\m\n}h_{\m\n}\right)\quad,\quad h^{\m\n}=g^{\m\a}g^{\n\b}h_{\a\b}
%\ee
%The extra term in the Einstein field equation \eqref{EFEg} proportional to $g_4$ is
%\be
%T^{(g_4)}_{\mu\nu}=F^2\left(F^2 g_{\m\n}-8{F_{\m}}^\a F_{\nu\a}\right)
%\ee
%whose first order perturbation is
%\begin{align}
%\delta T^{(g_4)}_{\mu\nu}&=h_{\m\n}F^4+8F^2h_{\a\b}{F_\mu}^\b {F_\n}^\a+2g_{\m\n}F^2F^{\a\b}f_{\a\b}+8F^2({F^\a}_\m f_{\n\a}+{F^\a}_\n f_{\m\a})\nn\\
%&+2f_{\a\b}F^{\a\b}(g_{\m\n}F^2+8{F^\r}_\m F_{\n\r})
%\end{align}
%In \eqref{EFEf} the extra term proportional to $g_4$ is
%\be
%J^\n=8(  F^2\nabla_\m F^{\m\n}+F^{\m\n}\nabla_\m F^2)
%\ee
\FloatBarrier
\section{Conclusions and discussion}
\label{sec:conclusions}
In this work, we addressed a problem of finding scalar QNMs of Reissner-Nordstr\"om black hole geometry in the near extremal regime. We incorporated the EFT corrections to the Einstein-Maxwell theory, and we considered a black hole solution including the perturbative corrections required by the presence of higher derivative operators.

In the WKB viewpoint, since the effective potential of the scalar wave in the deformed RN black hole background exhibits an unstable light ring that allows for asymptotically circular geodesics, the spectrum of scalar QNMs shows the prompt ringdown modes. These modes for the astrophysical black holes are associated with the first signal produced by the newly born compact object formed after the merger. This initial sequence of waves is linked to the unstable photon sphere: a wave impinging from infinity is partially scattered by the potential barrier and then detected by an observer at infinity. These modes have been carefully analyzed and computed using the WKB and eikonal approximations, as well as the Leaver continuous fraction method, together with numerical integration techniques.

Another class of QNMs is constituted by the so-called Zero Damping Modes (ZDMs), which arise in the near-extremal regime. We obtained an analytic expression for the frequencies of the ZDMs at linear order in EFT corrections. These modes have only an imaginary part, which is proportional to $r_+ - r_-$, and therefore tends to vanish in the near-extremal limit. In astrophysical settings, these modes are connected to the phenomenon of superradiance of Kerr black holes \cite{Brito:2015oca,Bianchi:2021mft,Bianchi:2023rlt,Cipriani:2024ygw}. In these situations, ZDMs acquire a real part that coincides precisely with the superradiant frequency. This frequency represents the threshold that must be exceeded in order for waves reflected by the black hole to have an amplitude larger than that of the incident ones. This phenomenon constitutes the wave analogue of the Penrose process.

The presence of ZDMs can be in a tight connection with Aretakis instability \cite{Lucietti:2012xr,Hadar:2017ven,Chen:2025sim}, as it has been pointed out in \cite{Richartz:2017qep,Gelles:2025gxi}.   Although it is known that black holes in the Kerr-Newman family have wave equations that are stable under linear perturbations (meaning that modes with positive time growth do not appear in the QNM spectrum), the Aretakis instability affecting the extremal solution is unrelated to the mode analysis of the differential operator. Instead, it is associated with the branch points in the frequency domain of the Green function. These branch points are located at $\omega=m \Omega_h$, where $m$ is the azimuthal quantum number and $\Omega_h$ is the horizon frequency. In particular, \cite{Richartz:2017qep} demonstrates that the wave character of the mode solutions is lost near the event horizon for ZDMs, establishing that the corresponding frequency is never a (quasi)normal frequency.

Although we expect similar phenomena for extremal Kerr black holes, our results are mainly applicable to the microscopic charged black holes in the near extremal limit\footnote{In this paper, we discuss the case of microscopic RN black holes (though with masses larger than the Planck mass, in order to keep EFT a valid description of them), as the realistic black holes observed in Nature cannot have large values of the charge.}. We found that our computation is justified around the near extremal regime if the EFT expansion is still correct near the outer horizon for the chosen set of parameters. We checked our results with the use of the Leaver and numerical integration methods. 

The main observation following from our computation is the direct connection of the first correction to ZDM with the combination of EFT couplings known to be constrained by the WGC. 

A causality requirement for the gravitational EFTs is formulated in \cite{Melville:2024zjq} as a condition for imaginary parts of the QNMs frequencies. It prescribes that the EFT corrections to the damping rate should always be positive, i.e., the higher derivative operators should make QNMs more stable. We found that the QNMs causality condition for the scalar wave translates exactly to the statement derived from the WGC in \cite{Arkani-Hamed:2006emk}. This result unravels an interesting link between causality and the requirement that all black holes must be able to decay. 

It is important to mention that the attempts to obtain the most optimal constraints on $g_4$ and $h_4$ from scattering amplitudes in flat space meet difficulties related to the presence of the graviton pole in the forward limit. The results obtained so far outside the forward limit \cite{Henriksson:2022oeu} are still weaker than the black hole WGC, and allow for small violations of this statement \cite{Henriksson:2022oeu,CarrilloGonzalez:2023cbf,Knorr:2024yiu}. Interestingly, the setup of the scalar wave on top of a near extremal RN black hole allows us to obtain the positivity of the WGC combination $2 g_4+h_4$ from the causality constraint for QNMs. The tensor and vector QNMs recently computed in \cite{Boyce:2025fpr} are sensitive only to $g_4$ at the leading order in the extremality parameter. Remarkably, previous studies of the gravitational EFTs \cite{Melville:2024zjq} are also showing that the QNM causality condition aligns with the constraints from positivity bounds and predictions from string theory. 

\subsection*{Acknowledgements}
The authors are indebted to Calvin Y.-R. Chen for the feedback and illuminating discussions on the EFT validity for extremal black holes. GDR and AT are grateful to Ivano Basile, Massimo Bianchi, Donato Bini, Francesco Fucito, Scott Melville and Jos\'e Francisco Morales for careful reading of our draft, valuable comments, and suggestions. The work of AT was supported by the National Natural Science Foundation of China (NSFC) under Grant No. 1234710.

\bibliography{biblioRNdef}
\end{document}